\documentclass[preprintnumbers,nofootinbib,floats,amssymb,floatfix]{revtex4-2}
\usepackage{amsfonts}
\usepackage{amsmath}
\usepackage{graphicx}
\usepackage{amssymb}
\usepackage{hyperref}
\usepackage{amsxtra}
\usepackage{amstext}
\usepackage{orcidlink}
\usepackage{endnotes}
\hypersetup{colorlinks,linkcolor={magenta},citecolor={cyan},urlcolor={magenta}}
\begin{document}
\title{Faddeev–Jackiw Hamiltonian formulation for general exotic bi-gravity}
\author{Omar Rodr{\'i}guez-Tzompantzi\orcidlink{0000-0002-6307-7203}} \email{omar.tzompantzi@unison.mx}
\affiliation{Departamento de Investigaci\'on en F\'isica, Universidad de Sonora,
Apartado Postal 5-088, C.P. 83000, Hermosillo, Sonora, M\'exico.}
\begin{abstract}
General exotic bi-gravity, obtained in Ozkan et al. (Phys Rev Lett 123(3):031303, 2019), is a unitary parity-preserving model which describes two interacting spin-two fields in three-dimensional spacetime. Adopting a symplectic viewpoint, we investigate the dynamical structure of general exotic bi-gravity theory. In particular, by exploiting the properties of the corresponding pre-symplectic matrix and its associated zero-modes, we explicitly derive all constraints of the theory, including the integrability conditions and scalar relationships between all the parameters and fields defining the  model. Then, as an application, these scalar relationships are used for studying the anti-de Sitter background. After that, we derive the gauge transformations for the dynamical variables from the structure of the remaining zero-modes, meaning that such zero-modes are indeed the generators of the gauge symmetry of the theory. Finally, by switching off one of the four coupling constants $\beta_{n}$ and assuming the invertibility of some of the dreibeins, we find that the general exotic bi-gravity theory has two physical degrees of freedom.
\end{abstract}
\preprint{}
\maketitle
\section{Introduction}
Three-dimensional (3D) gauge theories  have long been of significant interest in theoretical physics  because they often provide a relatively easy playing ground to test ideas that are difficult to prove in actual four-dimensions. Moreover, the topology of odd-dimensional spacetime  allows the construction of gauge field models with novel and attractive features. The prime example for us are gravitational theories describing interactions of massless or  massive spin-two fields with other spin-two fields,  and whose action can be written in the first-order formulation where the Lorentz and the diffeomorphism invariances are manifest \cite{Hohm, Achucarro, Routh, Witten, Witten2, Horne}. Interestingly enough, the construction of this class of models must be such that, at the nonlinear level, each resulting model must be able to ensure the vanishing of additional degrees of freedom (DoF), some of which could behave as negative energy ghost modes \cite{ghots}, allowing us to guarantee the consistency and stability of the theory. This is particularly important since consistent interacting theories of spin-two fields have proven to be useful theoretical laboratories, not only for investigating fundamental issues of quantum gravity \cite{Carlip, Afshar,Carlip2,Carlip3,Karim,Deser3,Keszthelyi,Johansson,Oda1,Oda2}, black hole physics \cite{banados,Strominger,Zhang,holes,Ghodsi}, and holography \cite{Paul,Joydeep} but also for studying the gapped dynamics of spin-two excitations observed in certain condensed-matter systems \cite{Gromov, Gromov2}. Bearing this in mind, the search for interacting theories for spin-two fields that are potentially free of ghost instabilities has been the subject of a series of interesting papers. The problem here is that the right number of physical DoF is not so obvious.

With respect to the DoF, the additional unphysical DoF in a given spin-2 fields model, and, in fact, in any other physical model, must be successfully eliminated by taking into account all the physical constraints and the gauge symmetry derived through a Hamiltonian analysis of it \cite{Dirac,Henneaux, Rothe,FJ}. In this sense, the Hamiltonian formulation of spin-two fields theories is undoubtedly a topic worth investigating, because it enables us to find several essential dynamical properties of the theory, such as  the constraints structure, the gauge symmetry and its generators, and the correct number of DoF. 

Regarding the action for a consistent interacting model of spin-2 fields in three-dimensions, it can be constructed  by using dreibeins and dualized spin-connections valued in the Lie algebra of the 3D Lorentz group \cite{Hohm,Achucarro,Witten,Witten2,Horne}. The simplest example is the Einstein-Cartan—also known as Palatini— action, of even parity, for pure Einstein gravity \cite{Achucarro, Witten}. A Hamiltonian analysis of such an action, with or without a cosmological constant, shows that such a theory is essentially a topological one, in the sense that its action has no DoF. In other words, there is no propagation of spin-two modes (gravitons) in three-dimensions. As a consequence,  the Newtonian interactions between point mass objects as well as gravitational waves are absent \cite{Deser,Hooft}. Nevertheless, the theory, with a negative cosmological constant, admits asymptotically AdS black hole solutions and has a dual description through AdS/CFT correspondence \cite{Paul,banados,Strominger}. Interestingly, when we supplement the Palatini action with the parity-violating Lorentz-Chern-Simons (LCS) term of  3D conform  gravity \cite{Horne}, its physical content becomes richer. The resulting theory,  dubbed as Topologically Massive Gravity (TMG) \cite{Deser1,Deser2,Hohm}, breaks parity symmetry with a new mass scale parameter. Despite some challenges \cite{Grumiller,Carlip4, Mu}, the Hamiltonian formulation of the TMG theory  shows that it is free of ghost instabilities.  At the same time, it does exhibit emergent one physical DoF at the nonlinear level \cite{Cvetkovic}, which, for linear perturbations around  a background, can be recognized as a single propagating massive spin-two mode.   A plausible parity-even extension of TMG has been constructed and analyzed in \cite{Bergshoeff2,Hohm} under the name of New Massive Gravity (NMG). The Hamiltonian formulation, in this case, reveals that, at the nonlinear level, NMG has no ghosts, and therefore it describes two physical DoF in total, corresponding to a doublet of spin-two modes of equal mass \cite{Blagojevic}. At the linearized level, NMG turns out to be equivalent to the 3D version of the first massive spin-two field theory, proposed long ago by Fierz and Pauli \cite{Pauli}. Both models have been the base to construct new parity-even and parity-odd models free from ghosts but with physical DoF realized as massive spin-two modes \cite{Bergshoeff3,Bergshoeff4,GMMG,Wout}.

On the other hand, one can find theories of two massive spin-two fields  in three-dimensions,  denoted as bi-gravity models, which interact with each other through non-derivative interaction terms.  The first example of such theories is the Zwei-Dreibein Gravity (ZDG) model \cite{ZDG,Hamid2} which has as a limiting case to NMG, and whose action consists of two copies of the parity-even Palatini action glued together by a cubic potential involving the two dreibeins. At the nonlinear level, a first Hamiltonian analysis found that, for generic parameters, ZDG contains three DoF corresponding to two spin-two modes and one ghost \cite{ZDG2}, even though it is ghost-free in a linearised approximation. Nevertheless, this instability was cured by identifying an additional constraint, which can only be derived from integrability conditions if we demand the invertibility of a linear combination of two dreibeins. In this environment, ZDG is then free of ghosts and hence has only two physical DoF, corresponding to two interacting spin-two modes of equal mass, as expected. The addition of an LCS term to the ZDG ($\beta_{2}=0$) action, leads to the parity-violating general ZDG model, which has shown to be ghost-free and has the same number of spin-2 modes as ZDG, but of different masses \cite{Routh}. Both models can be viewed as a 3D version of the Hassan and Rosen model for bi-gravity \cite{Hassan}. 

In the present work, we are predominantly interested in the study of the recently proposed exotic version of ZDG theory, dubbed as general exotic bi-gravity in Ref. \cite{GEMG}. It preserves parity and describes a unitary theory based on the same set of fields as the ZDG model; namely, two dreibeins and two spin-connections. Moreover, it is exotic in the sense that its action is obtained from two copies of parity-odd exotic Einstein-Cartan actions, in which the role of the curvature terms is played by LCS terms. These two actions are coupled through one derivative interaction term which is odd under parity. Remarkably, a Hamiltonian analysis of general exotic bi-gravity theory has not been performed in detail in the literature yet. This is what we will undertake in this paper.

Before plunging into the analysis of the dynamics of general exotic bi-gravity, it should be noted that the Hamiltonian analysis of all the models of spin-two fields discussed up to this point, is based on Dirac's canonical formalism. According to Dirac's procedure \cite{Dirac,Henneaux}, the primary constraints emerge when the canonical momenta are computed. Then the consistency under time evolution of the primary constraints might lead to secondary constraints, and so on. At every step, if the Poisson bracket of a constraint with the total Hamiltonian is not either identically zero or expressed in terms of the constraints themselves, it means either a restriction on the Lagrangian multipliers, or yet more constraints. After identifying the whole set of constraints, they must be cataloged as first- or second-class ones by taking into account the Poisson brackets of all constraints with each other. Only after classifying all the constraints, does one do a proper accounting of physical DoF; the common lore says that each first-class constraint kills two DoF and each second-class constraint kills one. Besides that, according to the Dirac's conjecture, all first-class constraints are generators of the gauge transformations, and consequently the gauge transformation for an arbitrary dynamical variable can be generated by its Poisson bracket with a linear combination of first-class constraints \cite{Henneaux}. However, after reading the literature, we have noted that the Hamiltonian analysis \`a la Dirac for these kinds of theories is very tedious and involves heavy algebraic manipulations. To be more precise, it is still difficult to identify the explicit form of all constraints and check the consistency of the combined set of primary and secondary constraints, which in turn makes it difficult  to  classify them correctly, which can hide the dynamical structure of these theories.

Taking the previously mentioned observations as inspiration points, and noting the hard task of evaluating all commutation relations among all constraints, in the current paper we will employ an alternative approach to study the dynamical structure of general exotic bi-gravity. More precisely, we will apply the protocol of Faddeev–Jackiw Hamiltonian formulation \cite{FJ,Omar,Montani1,Montani2}. Such a formalism is geometrically well-motivated and is based on the symplectic structure of the space phase. As we will see, this canonical symplectic approach offers several advantages, such as not having to define conjugate momenta for those velocities that appear linearly in the action, not having to use the Poisson brackets algebra, not having to distinguish between different types of constraints, and not having to rely on Dirac’s conjecture. Hence, we can expect that the algebraic manipulations needed in the Hamiltonian treatment of general exotic bi-gravity could be shortened. Indeed, we will show that the structure of all physical constraints, the gauge symmetry and its generators, can be obtained by simply analyzing  the properties of the pre-symplectic two-form matrix and its corresponding zero-modes. In doing so,  we will be able to determine the number of physical DoF propagated in the theory by taking into account all physical constraints and the gauge symmetry obtained from the Faddeev–Jackiw Hamiltonian description. In an earlier work \cite{Omar}, we showed that  the Faddeev-Jackiw approach is conceptually simpler and algebraically easier to implement in theories of interacting spin-2 fields. 

This manuscript is organized as follows. In Sec. \ref{Introduction}, we will review the action principle of general exotic bi-gravity and perform the $2+1$  decomposition of the action to identify the canonical variables that make up the pre-symplectic matrix. In Sec. \ref{framework}, we will employ the Faddeev-Jackiw symplectic procedure to derive the whole set of physical constraints. In Sec. \ref{Symmetries}, we will determine the gauge transformations for the dynamical variables using the zero-modes of the pre-symplectic matrix. In Sec. \ref{Diffeo}, we will cover the diffeomorphism symmetry by mapping the gauge parameters appropriately. In Sec. \ref{DoF}, we will count the number of the physical degrees of freedom by taking into account all the physical constraints and the gauge symmetry. In Sec. \ref{conclusions}. we will conclude our results.

We conclude with some comments on notation. Our convention for the indices throughout this article is such that Greek indices $\mu,\nu,\dotsc,$ will be spacetime indices; Latin indices $i,j,\dotsc,$ will be space indices; and Latin indices $a,b,\dotsc$ will be Lorentz indices.

\section{General exotic bi-gravity}
\label{Introduction}
\subsection{The action}
We begin with a brief review of the action for general exotic bi-gravity theory (see \cite{GEMG} for more details). Let us first assume that the space-time $\mathcal{M}$ is an orientable 3D manifold equipped with a Levi-Civita tensor $\varepsilon^{\mu\nu\rho}$. Then, on $\mathcal{M}$, we take the same set of fields as the ZDG model \cite{ZDG}; namely, a pair of dreibein one-forms $e^{a}=e^{a}_{\mu}dx^{\mu}$ and $l^{a}=l^{a}_{\mu}dx^{\mu}$, and a pair of dualised spin-connection one-forms $w_{a}=\epsilon_{abc}w{_{\mu}}^{bc}dx^{\mu}$ and $A_{a}=\epsilon_{abc}A{_{\mu}}^{bc}dx^{\mu}$ valued on the adjoint representation of the Lie group $SO(2,1)$, so that, it admits an invariant totally anti-symmetric tensor $\epsilon^{abc}$. The action of general exotic bi-gravity is given by the sum of four parts \cite{GEMG}:
\begin{eqnarray}
S[e^{a}_{\alpha},w^{a}_{\alpha},l^{a}_{\alpha},A^{a}_{\alpha}]&=&\int\limits_{\mathcal{M}} d^{3}x\Bigl(\mathcal{L}^{1}[e^{a}_{\alpha},w^{a}_{\alpha}]+\mathcal{L}^{2}[l^{a}_{\alpha},A^{a}_{\alpha}]+\mathcal{L}^{\text{int}}[e^{a}_{\alpha},w^{a}_{\alpha},l^{a}_{\alpha},A^{a}_{\alpha}]+\mathcal{L}^{\text{pot}}[e^{a}_{\alpha},l^{a}_{\alpha}]\Bigr). \label{principle}
\end{eqnarray}
The  main building blocks  of this action are two copies  of ``exotic'' Einstein-Cartan Lagrangians, constructed from the parity-violating Lorentz-Chern-Simons terms plus torsion terms, each being independently invariant under diffeomorphisms as well as internal Lorentz transformations. So, they read:
\begin{eqnarray}
\mathcal{L}^{1}[e^{a}_{\alpha},w^{a}_{\alpha}]&=&\frac{1}{2}\alpha_{1}\varepsilon^{\mu\nu\gamma}\left(w_{a\mu}\partial_{\nu}w_{\gamma}^{a}+\frac{1}{3}\epsilon^{abc}w_{a\mu}w_{b\nu}w_{c\gamma}\right)+\beta_{1}\varepsilon^{\mu\nu\gamma}e_{a\mu}D_{\nu}e^{a}_{\gamma},\label{C-S1}\\
\mathcal{L}^{2}[l^{a}_{\alpha},A^{a}_{\alpha}]&=&\frac{1}{2}\alpha_{2}\varepsilon^{\mu\nu\gamma}\left(A_{a\mu}\partial_{\nu}A_{\gamma}^{a}+\frac{1}{3}\epsilon^{abc}A_{a\mu}A_{b\nu}A_{c\gamma}\right)+\beta_{2}\varepsilon^{\mu\nu\gamma}l_{a\mu}\nabla_{\nu}l^{a}_{\gamma}\label{C-S2},
\end{eqnarray}
where $\alpha_{1}$, $\alpha_{2}$, $\beta_{1}$, and $\beta_{2}$, are dimensionful parameters.  In particular, we have two types of covariant derivative, acting on Lorentz indices, by mean of the spin-connections $w_{\alpha}^{a}$ and $A_{\alpha}^{a}$, respectively,
 \begin{eqnarray}
D_{\nu}V^{a}_{\gamma}&=&\partial_{\nu}V_{\gamma}^{a}+\epsilon^{abc}w_{b\nu}V_{c\gamma},\\
\nabla_{\nu} V^{a}_{\gamma}&=&\partial_{\nu}V_{\gamma}^{a}+\epsilon^{abc}A_{b\nu}V_{c\gamma},
\end{eqnarray}
where $\partial$ is a fiducial derivative operator. 
The  coupling between (\ref{C-S1}) and (\ref{C-S2}) is given by the interaction term
\begin{equation}
\mathcal{L}^{\text{\text{int}}}[e^{a}_{\alpha},w^{a}_{\alpha},l^{a}_{\alpha},A^{a}_{\alpha}]=\varepsilon^{\mu\nu\gamma}\left(\beta_{3}e_{a\mu}\nabla_{\nu}e^{a}_{\gamma}+\beta_{4}l_{a\mu}D_{\nu}l^{a}_{\gamma}\right),
\end{equation}
where $\beta_{3}$ and $\beta_{4}$ are new coupling constants. It was discussed in Ref. \cite{GEMG} that $\mathcal{L}^{1}$, $\mathcal{L}^{2}$, and $\mathcal{L}^{int}$, are all parity-odd. This implies that at the lineal level the kinetic terms of the two massive modes have the opposite signs, meaning one of them must be a Boulware-Deser mode, which violates the unitarity of the theory. Hence, in order to recover unitarity, it is necessary to add the following parity-even potential:
\begin{equation}
 \quad\mathcal{L}^{\text{pot}}[e^{a}_{\alpha},l^{a}_{\alpha}]=\varepsilon^{\mu\nu\gamma}\epsilon_{abc}\left(\gamma_{1}e_{\mu}^{a}e_{\nu}^{b}e_{\gamma}^{c}+\gamma_{2}l_{\mu}^{a}l_{\nu}^{b}l_{\gamma}^{c}\right),
\end{equation}
which carries two more parameters $\gamma_{1}$ and $\gamma_{2}$. Note that the curvature term is not present in the action (\ref{principle}), which is a special feature of the exotic model.
 
 In the next subsection we will start the canonical analysis for general exotic bi-gravity theory.


\subsection{The (2+1) decomposition of spacetime} 
In order to find the canonical form of the action (\ref{principle}) it is useful to decompose spacetime into space plus time. Let us assume not only that the spacetime manifold $\mathcal{M}$ has the topology $\mathcal{M}\cong\Re\times \Sigma$, where $\Sigma$ is a space-like hypersurface without boundaries parametrized by the coordinates $x^{i}$, while $\Re$ defines the temporal direction parametrized by $x^{0}$, but also that simultaneous proper $(2+1)$ decompositions exist for the pair of dreibeins as well as for the two connections. Splitting all spacetime indices into time and spatial components, the action (\ref{principle}) can be put, up to a boundary term, into the canonical form:
\begin{eqnarray}
S&=&\int\limits_{\Re\times\Sigma}dtd^{2}x\Bigl[\varepsilon^{0ij}\left(\frac{1}{2}\alpha_{1}\dot{w}_{ai}w^{a}_{j}+\beta_{13}^{+}\dot{e}_{ai}e^{a}_{j}+\beta_{24}^{+}\dot{l}_{ai}l^{a}_{j}+\frac{1}{2}\alpha_{2}\dot{A}_{ai}A^{a}_{j}\right)+ w_{a0}\Phi^{a}_{1}+e_{a0}\Phi^{a}_{2}+A_{a0}\Phi^{a}_{3}+l_{a0}\Phi^{a}_{4}\Bigr].\label{action}
\end{eqnarray}
The functions $\Phi^{a}_{1}$, $\Phi^{a}_{2}$, $\Phi^{a}_{3}$ and $\Phi^{a}_{4}$ appearing above are given by,
\begin{eqnarray}
\Phi^{a}_{1}&=&\varepsilon^{0ij}\Bigl(\alpha_{1}R_{ij}^{a}+\epsilon^{abc}\left(\beta_{1}e_{bi}e_{cj}+\beta_{4}l_{bi}l_{cj}\right)\Bigl),\label{P1}\\
\Phi^{a}_{2}&=&\varepsilon^{0ij}\left(2\beta_{1}D_{i}e^{a}_{j}+2\beta_{3}\nabla_{i}e^{a}_{j}+3\gamma_{1}\epsilon^{abc}e_{bi}e_{cj}\right),\label{P2}\\
\Phi^{a}_{3}&=&\varepsilon^{0ij}\left(\alpha_{2}F_{ij}^{a}+\epsilon^{abc}\left(\beta_{2}l_{bi}l_{cj}+\beta_{3}e_{bi}e_{cj}\right)\right),\label{P3}\\
\Phi^{a}_{4}&=&\varepsilon^{0ij}\left(2\beta_{2}\nabla_{i}l^{a}_{j}+2\beta_{4}D_{i}l^{a}_{j}+3\gamma_{2}\epsilon^{abc}l_{bi}l_{cj}\right),\label{P4}
\end{eqnarray}
where $\beta_{13}^{+}=\left(\beta_{1}+\beta_{3}\right)$ and $\beta_{24}^{+}=\left(\beta_{2}+\beta_{4}\right)$.
The spatial components of the curvatures and the covariant derivatives have the following form:
\begin{eqnarray}
R_{ij}^{a}&=&\partial_{i}w_{j}^{a}+\frac{1}{2}\epsilon^{abc}w_{bi}w_{cj},\\
 F_{ij}^{a}&=&\partial_{i}A_{j}^{a}+\frac{1}{2}\epsilon^{abc}A_{bi}A_{cj},\\
D_{i}V_{j}^{a}&=&\partial_{i}V^{a}_{j}+\epsilon^{abc}w_{bi}V_{cj},\\
\nabla_{i}V_{j}^{a}&=&\partial_{i}V^{a}_{j}+\epsilon^{abc}A_{bi}V_{cj},
\end{eqnarray}
with $V_{j}^{a}\in(e_{j}^{a},l_{j}^{a})$. Now, it is easy to see that the above action (\ref{action}) is first order in time derivatives. For our purpose it will be easier to work with an equivalent formulation of (\ref{action}) \cite{FJ},
\begin{equation}
S[\xi_{I}]=\int_{\Re\times \Sigma}\left(a^{I}[\xi]{}\dot{\xi}_{I}-\mathcal{V}[\xi]\right)dt\,d^{2}x,\label{action2}
\end{equation}
where the functionals $\xi_{I}$ stands for the set of field canonical variables describing the theory, and $a^{I}$ stands for the components of the canonical one-form $a(\xi)=a_{I}(\xi)d\xi^{I}$.  The general compound index $I$  runs in the different ranges of the complete set of variables. In our case, the original set of field variables $\xi^{I}$ is given by the set of dynamical fields $(w_{a}^{i},e_{a}^{i},A_{a}^{i}, l_{a}^{i})$ plus a set of Lagrangian multipliers $(w_{a}^{0},e_{a}^{0},A_{a}^{0}, l_{a}^{0})$:
\begin{equation}
\xi_{I}=\left(w_{a0},w_{ai},e_{a0},e_{ai},A_{a0},A_{ai},l_{a0},l_{ai}\right).\label{var}
\end{equation}
Then the components of the corresponding canonical one-form are:
\begin{equation}
a^{I}=\varepsilon^{0im}\left(0,\frac{1}{2}\alpha_{1}w_{m}^{a},0,\beta^{+}_{13}e_{m}^{a},0,\frac{1}{2}\alpha_{2}A_{m}^{a},0,\beta^{+}_{24}l_{m}^{a}\right).\label{1form}
\end{equation}
The functional $\mathcal{V}[\xi]$ is the symplectic potential, and since it does not contain time derivatives of the variables $\xi_{I}$, $\mathcal{V}[\xi]$ is defined as being the Hamiltonian functional $\mathcal{H}$; and so
\begin{equation}
\mathcal{H}[\xi]=-w_{a0}\Phi^{a}_{1}-e_{a0}\Phi^{a}_{2}-A_{a0}\Phi^{a}_{3}-l_{a0}\Phi^{a}_{4}.\label{potential}
\end{equation}
Now, the variation of the action (\ref{action2}) under a arbitrary variation $\delta\xi_{I}$ yields the following field equations:
\begin{equation}
\int d^{2}x\left(\mathcal{F}^{IJ}[x,y]\dot{\xi}_{J}[y]-\frac{\delta }{\delta\xi_{I}[y]}\mathcal{H}[x]\right)=0.\label{E-M}
\end{equation}
The elements of the called  symplectic matrix $\mathcal{F}_{IJ}(\xi)$ are the components of the symplectic two-form $\mathcal{F}[\xi]=d a[\xi]$ and are given in terms of a generalized curl,
\begin{equation}
\mathcal{F}_{IJ}[x,y]=\frac{\delta a_{J}[y]}{\delta\xi^{I}[x]}-\frac{\delta a_{I}[x]}{\delta\xi^{J}[y]}.\label{F}
\end{equation}
Clearly, $\mathcal{F}_{IJ}$ is an anti-symmetrical matrix which completely characterizes the dynamics of the theory. In our case the symplectic matrix $\mathcal{F}_{IJ}[x,y]$, whose matrix elements were written in (\ref{F}), takes the form,
\begin{equation}
\mathcal{F}_{IJ}[x,y]=\varepsilon^{0ji}
\eta_{ab}\left(
  \begin{array}{cccccccc}
 0   & 0  &  0   &  0     &  0  &  0&  0  &  0 	 	 \\                                                                        
 0 & \alpha_{1}   & 0   &  0 &   0   & 0 &  0  &  0\\                                                                   
0  & 0  &  0 &0  & 0 	& 0 	 &  0  &  0\\
0   &  0   &  0     &  2\beta^{+}_{13}    &   0   &  0&  0  &  0 	\\
    0   & 0   &0    & 0  & 0	 &  0&  0  &  0 \\
0  & 0  &  0 &0  & 0 	& \alpha_{2} 	 &  0  &  0\\
0  & 0  &  0 &0  & 0 	& 0 	 &  0  &  0\\
0   &  0  & 0  & 0   &  0 	&  0&  0  &  2\beta^{+}_{24}
 \end{array}
\right)\delta^{2}(x-y).\label{sym1}
\end{equation}
It is easy to convince that the matrix $\mathcal{F}_{IJ}$ is singular, since $det\, \mathcal{F}_{IJ}=0$. Thus, strictly speaking, we cannot identify $\mathcal{F}_{IJ}$ as a symplectic matrix. Nevertheless, this can be defined as a pre-symplectic matrix. This feature reveals that our model is  degenerated since there are more degrees of freedom in its field equations (\ref{E-M}) than physical degrees of freedom in the theory. To avoid this inconsistency, the theory must contain constraints that reduce the number of independent degrees of freedom and maintain the consistency of the field equations in the model. Below we show how to obtain all constraints of our model in the symplectic framework.

\section{ The nature of constraints in the symplectic framework}
\label{framework}
In this section deals with the derivations of the whole set of constraints on the dynamical variables, which must remove the additional unphysical degrees of freedom. 
\subsection{Primary constraints}
In view of  $det\, \mathcal{F}_{IJ}=0$,  the pre-symplectic matrix $\mathcal{F}_{IJ}$  necessarily has some independent zero-modes. They read,
\begin{eqnarray}
v_{1}^{I}&=& \left (v^{w_{0}},0 ,0,0, 0,0,0,0 \right),\\
v_{2}^{I}&=& \left(0,0,v^{e_{0}},0,0,0,0,0\right),\\
v_{3}^{I}&=& \left(0,0,0,0,v^{A_{0}},0,0,0\right),\\
v_{4}^{I}&=& \left(0,0,0,0,0,0,v^{l_{0}},0\right),
\end{eqnarray}
where $v^{w_{0}}$, $v^{e_{0}}$, $v^{A_{0}}$, $v^{l_{0}}$ are arbitrary functionals. On the othe hand,  using the Hamiltonian (\ref{potential}) and the canonical variables (\ref{var}),  the column matrix $(\delta \mathcal{H} /\delta\xi_{I})$ in Eq. (\ref{E-M}) is given by,
\begin{equation}
\frac{\delta \mathcal{H}[x]}{\delta\xi_{I}[y]}=
\left(
  \begin{array}{c}
-\Phi_{1}^{a} \\
\alpha_{1}\varepsilon^{0ij}w_{b0}\mathbf{W}^{\mathbf{x}ab}_{j}+2\beta_{1}\varepsilon^{0ij}e_{b0}\mathbf{E}^{ab}_{j}+2\beta_{4}\varepsilon^{0ij}l_{b0}\mathbf{L}^{ab}_{j}\\ 
-\Phi_{2}^{a} \\
2\beta_{1}\varepsilon^{0ij}w_{b0}\mathbf{E}^{ab}_{j}+2\beta_{1}\varepsilon^{0ij}e_{b0}\mathbf{W}^{\mathbf{x}ab}_{j}+2\beta_{3}\varepsilon^{0ij}e_{b0}\mathbf{A}^{\mathbf{x}ab}_{j}+6\gamma_{1}\varepsilon^{0ij}e_{b0}\mathbf{E}^{ab}_{j}+2\beta_{3}\varepsilon^{0ij}A_{b0}\mathbf{E}^{ab}_{j}\\
-\Phi_{3}^{a} \\
\alpha_{2}\varepsilon^{0ij}A_{b0}\mathbf{A}^{\mathbf{x}ab}_{j}+2\beta_{2}\varepsilon^{0ij}l_{b0}\mathbf{L}^{ab}_{j}+2\beta_{3}\varepsilon^{0ij}e_{b0}\mathbf{E}^{ab}_{j} \\
-\Phi_{4}^{a} \\
2\beta_{2}\varepsilon^{0ij}A_{b0}\mathbf{L}^{ab}_{j}+2\beta_{2}\varepsilon^{0ij}l_{b0}\mathbf{A}^{\mathbf{x}ab}_{j}+2\beta_{4}\varepsilon^{0ij}l_{b0}\mathbf{W}^{\mathbf{x}ab}_{j}+6\gamma_{2}\varepsilon^{0ij}l_{b0}\mathbf{L}^{ab}_{j}+2\beta_{4}\varepsilon^{0ij}w_{b0}\mathbf{L}^{ab}_{j}
 \end{array}
\right)
\delta^{2}(x-y),\label{Z1}
\end{equation}
where we have abbreviated 
$\mathbf{W}^{\textbf{y}ab}_{m}=\partial_{m}^{\textbf{y}}\eta^{ab}+\epsilon^{abc}w_{cm}$,
$\mathbf{E}^{ab}_{m}=\epsilon^{abc}e_{cm}$,
$\mathbf{A}^{\textbf{y}ab}_{m}=\partial_{m}^{\textbf{y}}\eta^{ab}+\epsilon^{abc}A_{cm}$,
$\mathbf{L}^{ab}_{m}=\epsilon^{abc}l_{cm}$.

Now, taking into account that by definition each zero-mode of $\mathcal{F}_{IJ}$, denoted by $v^{I}$, satisfies the equation $\int v^{I}\mathcal{F}_{IJ} d^{2}x=0$, it follows that the contraction of field equations (\ref{E-M}) with zero-modes of the singular matrix (\ref{sym1}) leads to the following constraint equations:
\begin{eqnarray}
\int v{_{1}}^{I}[y]d^{2}y\int\frac{\delta\mathcal{H}[x]}{\delta\xi^{I}[y]} d^{2}x  &=&v^{w_{0}}\Phi_{1}^{a}=0,\\
\int v{_{2}}^{I}[y]d^{2}y\int\frac{\delta\mathcal{H}[x]}{\delta\xi^{I}[y]} d^{2}x  &=&v^{e_{0}}\Phi_{2}^{a}=0, \\
\int v{_{3}}^{I}[y]d^{2}y\int\frac{\delta\mathcal{H}[x]}{\delta\xi^{I}[y]} d^{2}x  &=&v^{A_{0}}\Phi_{3}^{a}=0,\\
\int v{_{4}}^{I}[y]d^{2}y\int\frac{\delta\mathcal{H}[x]}{\delta\xi^{I}[y]} d^{2}x  &=&v^{l_{0}}\Phi_{4}^{a}=0.
\end{eqnarray}
Because $v^{w_{0}}$, $v^{e_{0}}$, $v^{A_{0}}$ and $v^{l_{0}}$ are arbitrary functions, we find the following  twelve primary constraints:
\begin{eqnarray}
\Phi_{1}^{a}=0,\quad\Phi_{2}^{a}=0,\quad\Phi_{3}^{a}=0,\quad \Phi_{4}^{a}=0.\label{C-P2}
\end{eqnarray}
The explicit expressions for $\Phi_{1}^{a},\Phi_{2}^{a},\Phi_{3}^{a}, \Phi_{4}^{a}$ were previously defined in Eqs. (\ref{P1})-(\ref{P4}). A few comments here are in order. First, the above provides evidence that the zero-modes of the pre-symplectic matrix $\mathcal{F}_{IJ}$ generate constraints. Second, the presence of the primary constraints (\ref{C-P2}) implies the existence of a surface in the phase space, dubbed as primary constraint surface, to which physical field configurations are restricted. This means that the dynamics of our system does not take place in the phase space $\Gamma$ coordinatized by dynamical variables $(w_{a}^{i},e_{a}^{i},A_{a}^{i}, l_{a}^{i})$, but rather on the submanifold $\Gamma_{P}\subset\Gamma$ defined by all primary constraints $\Phi_{1,2,3,4}^{a}$.  Third, all the primary constraints have arisen directly from the projections of the field equations (\ref{E-M}), meaning they are valid at all times; i.e. $\dot{\Phi}_{1,2,3,4}^{a}=d\Phi_{1,2,3,4}^{a}/dt=0$ hold on $\Gamma_{P}$. In this way, $\dot{\Phi}_{1,2,3,4}^{a}|_{\Gamma_{P}}=0$ are guaranteed to hold and need not be imposed as extra conditions. Having said that, we shall combine the fields equations (\ref{E-M}) with the fact that $\dot{\Phi}_{1,2,3,4}^{a}\mid_{\Gamma_{P}}=0$ are automatically satisfied to explore whether there are more new constraints, as described below.
\subsection{Secondary constraints and integrability conditions}
\label{IntCondASec}
Let us now note that all primary constraints only depend on canonical variables $\xi_{I}$, such that the conditions $\dot{\Phi}_{1,2,3,4}^{a}=0$ can be written as:
\begin{equation}
\dot{\Xi}_{M}=\int d^{2}x\frac{\delta \Xi_{M}}{\delta \xi^{J}}\dot{\xi}^{J}=0,\label{EqPrimary}
\end{equation}
with $\Xi_{M}\in\left(\Phi^{a}_{1},\Phi^{a}_{2},\Phi^{a}_{3},\Phi^{a}_{4}\right)$. Then, combining Eq. (\ref{EqPrimary}) with field equation (\ref{E-M}), we obtain the following new system of linear equations:
\begin{equation}
\int d^{2}x  \left(\mathcal{F}^{(1)}_{KJ}[x,y]\dot{\xi}^{J}[y]- \mathcal{Z}^{(1)}_{K}[x,y]\right)=0.\label{secondEM}
\end{equation}
This last expression must be evaluated on $\Gamma_{P}$ because the dynamics of our system take place now in the primary constraint surface. The matrices $\mathcal{F}^{(1)}_{KJ}$ and $ \mathcal{Z}^{(1)}_{K}$ are given by,
\begin{equation}
\mathcal{F}^{(1)}_{KJ}[\xi]=
\left(
\begin{array}{cc} 
\mathcal{F}_{IJ}[x,y] \\ 
\frac{\delta\Xi_{M}[x]}{\delta\xi^{J}[y]}
\end{array}
\right) \label{F1}\quad
\text{and}
\quad
 \mathcal{Z}^{(1)}_{K}[\xi]=
\left(
\begin{array}{ccc} 
\frac{\delta \mathcal{H}[x]}{\delta \xi^{I}[y]} \\ 
0\\
0\\
0\\
0
\end{array}
\right).
\end{equation}
The submatrices $\mathcal{F}_{IJ}$ and $(\delta \mathcal{H}/\delta\xi^{I})$ were constructed previously in (\ref{sym1}) and (\ref{Z1}), respectively.   
On the other hand, using the set of primary constraints $\Xi_{M}$  and the canonical variables (\ref{var}), the submatrix $\left(\delta\Xi_{M}/\delta\xi^{J}\right)$ in  (\ref{F1})  is,
\begin{equation}
\left(\frac{\delta \Xi_{M}}{\delta\xi^{J}}\right)=
2\left(
  \begin{array}{cccccccc}
0   & \frac{\alpha_{1}}{2}\mathbf{W}^{\textbf{y}ab}_{m} & 0  & \beta_{1}\mathbf{E}_{m}^{ab} &0 &0&0 &\beta_{4}\mathbf{L}_{m}^{ab}  \\
0  & \beta_{1}\mathbf{E}_{m}^{ab} & 0 &\beta_{1}\mathbf{W}_{m}^{\textbf{y}ab}+\beta_{3}\mathbf{A}_{m}^{\textbf{y}ab}+3\gamma_{1}\mathbf{E}_{m}^{ab} & 0 & \beta_{3}\mathbf{E}_{m}^{ab} &0 &0\\
0&0&0&\beta_{3}\mathbf{E}_{m}^{ab}&0&\frac{\alpha_{2}}{2}\mathbf{A}_{m}^{\textbf{y}ab}&0&\beta_{2}\mathbf{L}_{m}^{ab}\\
0&\beta_{4}\mathbf{L}_{m}^{ab}&0&0&0&\beta_{2}\mathbf{L}_{m}^{ab}&0&\beta_{2}\mathbf{A}_{m}^{\textbf{y}ab}+\beta_{4}\mathbf{W}_{m}^{\textbf{y}ab}+3\gamma_{2}\mathbf{L}_{m}^{ab}
 \end{array}
\right)
\varepsilon^{0jm}\delta^{2}(x-y)\label{A2}
\end{equation}
It is straightforward to verify that the new matrix $\mathcal{F}_{KJ}^{(1)}$ is also a singular one, and, therefore $\mathcal{F}_{KJ}^{(1)}$ has the following set of non-trivial zero-modes:
\begin{eqnarray}
v^{(1)K}_{1}&=& \left(0,\mathbf{W}^{\mathbf{y}ab}_{i}, 0, \frac{\beta_{1}}{\beta_{13}^{+}}\mathbf{E}^{ab}_{i}, 0, 0, 0, \frac{\beta_{4}}{\beta_{24}^{+}}\mathbf{L}^{ab}_{i},\eta^{ab},0,0,0 \right)\delta^{2}(x-y),\\
v^{(1)K}_{2}&=& \left(0,2\frac{\beta_{1}}{\alpha_{1}}\mathbf{E}^{ab}_{i},0, \frac{1}{\beta_{13}^{+}}\Bigl(\beta_{1}\mathbf{W}^{\mathbf{y}ab}_{i}+\beta_{3}\mathbf{A}^{\mathbf{y}ab}_{i}+3\gamma_{1}\mathbf{E}^{ab}_{i}\Bigr), 0, 2\frac{\beta_{3}}{\alpha_{2}}\mathbf{E}^{ab}_{i}, 0, 0,0,\eta^{ab},0,0 \right)\delta^{2}(x-y),\\
v^{(1)K}_{3}&=& \left(0,0,0,\frac{\beta_{3}}{\beta_{13}^{+}}\mathbf{E}^{ab}_{i}, 0, \mathbf{A}^{\mathbf{y}ab}_{i}, 0, \frac{\beta_{2}}{\beta_{24}^{+}}\mathbf{L}^{ab}_{i},0,0,\eta^{ab},0 \right)\delta^{2}(x-y)\\
v^{(1)K}_{4}&=& \left(0,2\frac{\beta_{4}}{\alpha_{1}}\mathbf{L}^{ab}_{i},0,0,0,2\frac{\beta_{2}}{\alpha_{2}}\mathbf{L}^{ab}_{i},0, \frac{1}{\beta_{24}^{+}}\left(\beta_{2}\mathbf{A}^{\mathbf{y}ab}_{i}+\beta_{4}\mathbf{W}^{\mathbf{y}ab}_{i}+3\gamma_{2}\mathbf{L}^{ab}_{i}\right), 0,  0, 0,\eta^{ab}\right)\delta^{2}(x-y).
\end{eqnarray} 
Given these zero modes, we can turn to the task of looking for new constraints. Indeed, by multiplying these zero-modes to the two sides of Eq. (\ref{secondEM}), we get, after some manipulations, the following constraint equations: (the integration symbols $\int$ have been omitted for simplicity):
\begin{eqnarray}
v^{(1)K}_{1}\mathcal{Z}_{K}^{(1)}&=&-\epsilon_{dbg}\left(\frac{\beta_{4}}{\beta_{24}^{+}}l^{b}_{0}\Phi^{g}_{4}+\frac{\beta_{1}}{\beta_{13}^{+}}e^{b}_{0}\Phi^{g}_{2}+w^{b}_{0}\Phi^{g}_{1}\right)+2\frac{\beta_{2}\beta_{4}}{\beta_{24}^{+}}\varepsilon^{\alpha\beta\gamma}l_{d\alpha}\left(w^{c}_{\beta}-A^{c}_{\beta}\right)l_{c\gamma}\nonumber\\
&&+2\frac{\beta_{1}\beta_{3}}{\beta_{13}^{+}}\varepsilon^{\alpha\beta\gamma}e_{d\alpha}\left(w^{c}_{\beta}-A^{c}_{\beta}\right)e_{c\gamma}=0,\\
v^{(1)K}_{2}\mathcal{Z}_{K}^{(1)}&=&-\epsilon_{dbg}\left(\frac{\beta_{1}}{\beta_{13}^{+}}w_{0}^{b}\Phi^{g}_{2}+\frac{\beta_{3}}{\beta_{13}^{+}}A_{0}^{b}\Phi^{g}_{2}+3\frac{\gamma_{1}}{\beta_{13}^{+}}e_{0}^{b}\Phi^{g}_{2}+2\frac{\beta_{1}}{\alpha_{1}}e_{0}^{b}\Phi^{g}_{1}+2\frac{\beta_{3}}{\alpha_{2}}e_{0}^{b}\Phi^{g}_{3}\right)\nonumber\\
&&+2\frac{\beta_{1}\beta_{3}}{\beta_{13}^{+}}\varepsilon^{\alpha\beta\mu}\left(w_{d\alpha}-A_{d\alpha}\right)e_{c\beta}\left(w^{c}_{\mu}-A^{c}_{\mu}\right)+4\varepsilon^{\alpha\beta\mu}\left(\frac{\beta_{1}\beta_{4}}{\alpha_{1}}+\frac{\beta_{3}\beta_{2}}{\alpha_{2}}\right)l_{d\alpha}e_{c\beta}l^{c}_{\mu}=0,\\
v^{(1)K}_{3}\mathcal{Z}_{K}^{(1)}&=&-\epsilon_{dbg}\left(\frac{\beta_{2}}{\beta_{24}^{+}}l^{b}_{0}\Phi^{g}_{4}+\frac{\beta_{3}}{\beta_{13}^{+}}e^{b}_{0}\Phi^{g}_{2}+A^{b}_{0}\Phi^{g}_{3}\right)-2\frac{\beta_{2}\beta_{4}}{\beta_{24}^{+}}\varepsilon^{\alpha\beta\gamma}l_{d\alpha}\left(w^{c}_{\beta}-A^{c}_{\beta}\right)l_{c\gamma}\nonumber\\
&&-2\frac{\beta_{1}\beta_{3}}{\beta_{13}^{+}}\varepsilon^{\alpha\beta\gamma}e_{d\alpha}\left(w^{c}_{\beta}-A^{c}_{\beta}\right)e_{c\gamma}=0,\\
v^{(1)K}_{4}\mathcal{Z}_{K}^{(1)}&=&-\epsilon_{dbg}\left(\frac{\beta_{4}}{\beta_{24}^{+}}w_{0}^{b}\Phi^{g}_{4}+\frac{\beta_{2}}{\beta_{24}^{+}}A_{0}^{b}\Phi^{g}_{4}+3\frac{\gamma_{2}}{\beta_{24}^{+}}l_{0}^{b}\Phi^{g}_{4}+2\frac{\beta_{2}}{\alpha_{2}}l_{0}^{b}\Phi^{g}_{3}+2\frac{\beta_{4}}{\alpha_{1}}l_{0}^{b}\Phi^{g}_{1}\right)\nonumber\\
&&+2\frac{\beta_{2}\beta_{4}}{\beta_{24}^{+}}\varepsilon^{\alpha\beta\mu}\left(w_{d\alpha}-A_{d\alpha}\right)l_{c\beta}\left(w^{c}_{\mu}-A^{c}_{\mu}\right)+4\varepsilon^{\alpha\beta\mu}\left(\frac{\beta_{1}\beta_{4}}{\alpha_{1}}+\frac{\beta_{3}\beta_{2}}{\alpha_{2}}\right)e_{d\alpha}l_{c\beta}e^{c}_{\mu}=0.
\end{eqnarray}
So after evaluating these resulting expressions on $\Gamma_{P}$, we obtain the following set of algebraic equations:
\begin{eqnarray}
 \mathcal{X}_{d}&=&\frac{\beta_{2}\beta_{4}}{\beta_{24}^{+}}\varepsilon^{\alpha\beta\gamma}l_{d\alpha}\left(w^{c}_{\beta}-A^{c}_{\beta}\right)l_{c\gamma} +\frac{\beta_{1}\beta_{3}}{\beta_{13}^{+}}\varepsilon^{\alpha\beta\gamma}e_{d\alpha}\left(w^{c}_{\beta}-A^{c}_{\beta}\right)e_{c\gamma}=0,\label{Int1}\\
 \mathcal{Y}_{d}&=&\frac{\beta_{1}\beta_{3}}{\beta_{13}^{+}}\varepsilon^{\alpha\beta\mu}\left(w_{d\alpha}-A_{d\alpha}\right)e_{c\beta}\left(w^{c}_{\mu}-A^{c}_{\mu}\right)+2\varepsilon^{\alpha\beta\mu}\left(\frac{\beta_{1}\beta_{4}}{\alpha_{1}}+\frac{\beta_{3}\beta_{2}}{\alpha_{2}}\right)l_{d\alpha}e_{c\beta}l^{c}_{\mu}=0,\label{Int2}\\
 \mathcal{Z}_{d}&=&\frac{\beta_{2}\beta_{4}}{\beta_{24}^{+}}\varepsilon^{\alpha\beta\mu}\left(w_{d\alpha}-A_{d\alpha}\right)l_{c\beta}\left(w^{c}_{\mu}-A^{c}_{\mu}\right)+2\varepsilon^{\alpha\beta\mu}\left(\frac{\beta_{1}\beta_{4}}{\alpha_{1}}+\frac{\beta_{3}\beta_{2}}{\alpha_{2}}\right)e_{d\alpha}l_{c\beta}e^{c}_{\mu}=0,\label{Int3}
\end{eqnarray}
where the uncontracted fields have a free Lorentz index. These resulting equations correspond to the integrability conditions\footnote{In Ref. \cite{GEMG}, it was argued that the integrability conditions exist, however, they were not derived.} for general exotic bi-gravity theory, which must be satisfied on the constraints surface $\Gamma_{P}$ and never leave it. Looking at such integrability conditions, we can easily notice that such conditions do not  yield any new constraint, since they will always mix dynamical variables with Lagrange multipliers. Nevertheless, we provide below two scenarios (i)-(ii) where new constraints could arise:

\textit{(i)} $\beta_{3}=0$. In such a case, $\beta_{1}$, $\beta_{2}$, and $\beta_{4}$,  are required to be generic,  and $l_{d}^{\mu}$ has an inverse. Then  from Eqs. (\ref{Int1}) and (\ref{Int2}) it follows that:
\begin{eqnarray}
\Theta^{\alpha}&=&\varepsilon^{\alpha\beta\mu}l_{c\beta}\left(w^{c}_{\mu}-A^{c}_{\mu}\right)=0,\label{sec1}\\
\Gamma^{\alpha}&=&\varepsilon^{\alpha\beta\mu}l_{c\beta}e^{c}_{\mu}=0,\label{sec2}
\end{eqnarray}
and thus $\mathcal{Z}_{d}$ (\ref{Int3}) is automatically fulfilled. The condition $\Gamma^{\alpha}$ is often called the symmetrization condition, which is of great importance in connecting the first-order formulation to the metric formulation of the theory. Now, the temporal components of these equations yield  two secondary constraints. Specifically, they take the form:
\begin{eqnarray}
\Theta^{0}&=&\varepsilon^{0ij}l_{ci}\left(w^{c}_{j}-A^{c}_{j}\right)=0,\\
\Gamma^{0}&=&\varepsilon^{0ij}e_{ci}l^{c}_{j}=0.
\end{eqnarray}
On the other hand,  the spatial component  of Eqs. (\ref{sec1})-(\ref{sec2}) mixes dynamical variables with Lagrangian multipliers, namely:
\begin{eqnarray}
\Theta^{i}&=&\varepsilon^{0ij}\left(l_{c0}\left(w^{c}_{j}-A^{c}_{j}\right)-l_{cj}\left(w^{c}_{0}-A^{c}_{0}\right)\right)=0,\label{case}\\
\Gamma^{i}&=&\varepsilon^{0ij}\left(e_{cj}l^{c}_{0}-e_{c0}l^{c}_{j}\right)=0.\label{case1}
\end{eqnarray}
This mean that $\Theta^{i}$ and $\Gamma^{i}$ are not constraints. Finally, it is important to mention that analogously to the condition on $\beta_{3}$, we can impose the condition $\beta_{1}=0$ instead of $\beta_{3}=0$ and obtain the same results.

\textit{(ii)} $\beta_{4}=0$. Now, $\beta_{1}$, $\beta_{2}$, $\beta_{3}$ are required to be generic parameters and that $e_{a}^{\mu}$ is invertible.  In doing so, from Eq. (\ref{Int1}) and (\ref{Int3}), we infer the following equations:
\begin{eqnarray}
\Psi^{\alpha}&=&\varepsilon^{\alpha\beta\mu}e_{c\beta}\left(w^{c}_{\mu}-A^{c}_{\mu}\right)=0,\label{case21}\\
\Gamma^{\alpha}&=&\varepsilon^{\alpha\beta\mu}l_{c\beta}e^{c}_{\mu}=0.\label{case22}
\end{eqnarray}
Using these equations, it is easily checked that the Eq. (\ref{Int2}) is satisfied identically. Now, the temporal component of these equations induces two secondary constraints,
\begin{eqnarray}
\Psi^{0}&=&\varepsilon^{0ij}e_{ci}\left(w^{c}_{j}-A^{c}_{j}\right)=0,\\
\Gamma^{0}&=&\varepsilon^{0ij}e_{ci}l^{c}_{j}=0,\label{case2}\nonumber
\end{eqnarray}
while the spatial component of (\ref{case21}) and (\ref{case22}) generate equations for the Lagrange multipliers, 
\begin{eqnarray}
\Psi^{i}&=&\varepsilon^{0ij}\left(e_{c0}\left(w^{c}_{j}-A^{c}_{j}\right)-e_{cj}\left(w^{c}_{0}-A^{c}_{0}\right)\right)=0,\\
\Gamma^{i}&=&\varepsilon^{0ij}\left(e_{cj}l^{c}_{0}-e_{c0}l^{c}_{j}\right)=0.\nonumber
\end{eqnarray}
Note that the same results can be achieved by imposing $\beta_{2}=0$ instead of $\beta_{4}=0$. 

We have demonstrated that the general exotic bi-gravity theory has two equivalent sectors with the same number of secondary constraints. The existence of the two secondary constraints, in each case, is related to the invertibility of only one dreibein and not to the invertibility of some linear combination of the two dreibeine, as in the case of ZDG model \cite{ZDG,ZDG2}.  In both cases, the primary and the corresponding secondary constraints define a submanifold $\Gamma_{S}\subset \Gamma$, called secondary constraint surface, where all constraints discovered until now, including the equations for the Lagrange multipliers, vanish. In the next section, we will show that these two secondary constraints are enough to remove all the unwanted DoF in the theory.

\subsection{Consistency of secondary constraints}
For the time evolution to be consistent, it is necessary to impose the stability conditions for the secondary constraints. Given that, in each sector, we now know the structure of the secondary constraints, instead of choosing a specific sector,  let us analyze the generic case where there will be three secondary constraints, keeping in mind that the specific cases \textit{(i)} and \textit{(ii)} may be addressed by taking the limit $\beta_{3}\rightarrow 0$, or $\beta_{4}\rightarrow 0$, respectively.

In the generic case, which is particularly interesting because it captures the whole theory described by  (\ref{action}), all the coupling constants $\beta_{1,2,3,4}$ are non-vanishing. In order to obtain  independent secondary constraints from the integrability conditions (\ref{Int1})-(\ref{Int3}), it is covenient to note that $\mathcal{Y}^{d}=0$ and $\mathcal{Z}^{d}=0$,  in Eqs. (\ref{Int2}) and (\ref{Int3}) respectively, is equivalent to
\begin{equation}
\Psi^{\alpha}=0,\quad\Theta^{\alpha}=0,\quad\Gamma^{\alpha}=0,
\end{equation}
which, in turn, ensure that $\mathcal{X}^{d}=0$ holds.  With this, we now have three independent secondary constraints,$\Psi^{0}$, $\Theta^{0}$, $\Gamma^{0}$, which have the same structure as in the case \textit{(i)} and \textit{(ii)}. On the other hand, the spatial part of $\Psi^{\alpha},\,\Theta^{\alpha},\,\Gamma^{\alpha}$ generate equations for the Lagrange multipliers. With regard to the stability conditions of above three secondary constraints, they can be written down as
\begin{equation}
\dot{\Xi}_{L}^{(1)}=\int d^{2}x\frac{\delta \Xi_{L}^{(1)}}{\delta \xi^{J}}\dot{\xi}^{J}=0,\label{F2}
\end{equation}
with $\Xi_{L}^{(1)}\in\left(\Theta^{0}, \Gamma^{0},\Phi^{0}\right)$. Together with (\ref{secondEM}), they form a new linear system,
\begin{equation}
\int d^{2}x \left(\mathcal{F}^{(2)}_{MJ}[x,y]\dot{\xi}^{J}[y]- \mathcal{Z}^{(2)}_{M}[x,y]\right)=0,\label{F3}
\end{equation}
where $\mathcal{F}^{(2)}_{MJ}$ and $\mathcal{Z}^{(2)}_{M}$  stand for
\begin{equation}
\mathcal{F}^{(2)}_{MJ}[\xi]=
\left(
\begin{array}{cc} 
\mathcal{F}_{IJ}^{(1)}[x,y] \\ 
\frac{\delta\Xi_{L}^{(1)}[x]}{\delta\xi^{J}[y]}
\end{array}
\right)\quad\text{and}\quad
\mathcal{Z}^{(2)}_{M}(\xi)=
\left(
\begin{array}{ccc} 
\mathcal{Z}^{(1)}_{K} \\ 
0\\
0\\
0
\end{array}
\right).
\end{equation}

Here the new submatrix $\left(\delta\Xi_{L}^{(1)}/\delta\xi^{J}\right)$ is given explicitly by
\begin{eqnarray}
\frac{\delta\Xi_{L}^{(1)}}{\delta\xi^{J}}&=&
\left(
  \begin{array}{ccccccccccc}
0&-l^{b}_{i}&0&0&0&l_{i}^{b}&0&\left(w_{i}^{b}-A_{i}^{b}\right)\\
0&0&0&l_{i}^{b}&0&0&0&-e_{i}^{b}\\
0&-e_{i}^{b}&0&\left(w_{i}^{b}-A_{i}^{b}\right)&0&e_{i}^{b}&0&0
 \end{array}
\right)\varepsilon^{0ji}\delta^{2}(x-y),
\end{eqnarray}
while $\mathcal{F}_{IJ}^{(1)}$ and $\mathcal{Z}^{(1)}_{K}$ are given in  (\ref{F1}).  Again, one can easily verify that the matrix $\mathcal{F}^{(2)}_{MJ}$ is also singular, and so, it contains the following  zero-modes:
\begin{eqnarray}
v_{1}^{(2)M}&=& \left (v_{1}^{(1)K},0,0,0 \right)\delta^{2}(x-y),\\
v_{2}^{(2)M}&=& \left(v_{2}^{(1)K},0,0,0\right)\delta^{2}(x-y),\\
v_{3}^{(2)M}&=& \left(v_{3}^{(1)K},0,0,0\right)\delta^{2}(x-y),\\
v_{4}^{(2)M}&=& \left(v_{4}^{(1)K},0,0,0\right)\delta^{2}(x-y),\\
v^{(2)M}_{5}&=& \Bigl(0,-\frac{1}{\alpha_{1}}l_{i}^{b}, 0, 0, 0, \frac{1}{\alpha_{2}}l_{i}^{b}, 0, \frac{1}{2}\frac{1}{\beta_{24}^{+}}(w_{i}^{b}-A_{i}^{b}),0,0,0,0,\eta^{ab},0,0 \Bigr)\delta^{2}(x-y),\\
v^{(2)M}_{6}&=& \Bigl(0,0, 0, \frac{1}{2}\frac{1}{\beta_{13}^{+}}l^{b}_{i}, 0, 0, 0, -\frac{1}{2}\frac{1}{\beta_{24}^{+}}e^{b}_{i},0,0,0,0,0,\eta^{ab},0 \Bigr)\delta^{2}(x-y),\\
v^{(2)M}_{7}&=&\Bigl(0,-\frac{1}{\alpha_{1}}e^{b}_{i},0,\frac{1}{2}\frac{1}{\beta_{13}^{+}}(w^{b}_{i}-A^{b}_{i}),0, \frac{1}{\alpha_{2}}e^{b}_{i},0,0,0,0,0,0,0,0,\eta_{ab}\Bigr)\delta^{2}(x-y).
\end{eqnarray}
If we multiply the first four zero-mode by Eq. (\ref{F3}), we will find the same secondary constraints obtained previously, whereas if we multiply the zero-modes $v^{(2)M}_{5}$, $v^{(2)M}_{6}$, and $v^{(2)M}_{7}$ by Eq. (\ref{F3}), and integrate by parts, we get the following constraint relations (the integration symbols $\int$ are omitted for simplicity):
\begin{eqnarray}
v^{(2)M}_{5}\mathcal{Z}_{M}^{(2)}&=&\partial_{i}\Theta^{i}+\frac{1}{2}\frac{1}{\beta_{24}^{+}}\left(w^{a}_{0}-A^{a}_{0}\right)\Phi_{4a}+l^{a}_{0}\left(\frac{1}{\alpha_{2}}\Phi_{3a}-\frac{1}{\alpha_{1}}\Phi_{1a}\right)\nonumber\\
&&+\frac{\beta_{2}-\beta_{4}}{2\beta_{24}^{+}}\varepsilon^{\alpha\beta\gamma}\epsilon_{abc}\left(w_{\alpha}^{a}-A_{\alpha}^{a}\right)l_{\beta}^{b}\left(w_{\gamma}^{c}-A_{\gamma}^{c}\right)-\frac{3}{2}\frac{\gamma_{2}}{\beta_{24}^{+}}\varepsilon^{\alpha\beta\gamma}\epsilon_{abc}\left(w_{\alpha}^{a}-A_{\alpha}^{a}\right)l_{\beta}^{b}l_{\gamma}^{c}\nonumber\\
&&+\left(\frac{\beta_{1}}{\alpha_{1}}-\frac{\beta_{3}}{\alpha_{2}}\right)\varepsilon^{\alpha\beta\gamma}\epsilon_{abc}l_{\alpha}^{a}e_{\beta}^{b}e_{\gamma}^{c}+\left(\frac{\beta_{4}}{\alpha_{1}}-\frac{\beta_{2}}{\alpha_{2}}\right)\varepsilon^{\alpha\beta\gamma}\epsilon_{abc}l_{\alpha}^{a}l_{\beta}^{b}l_{\gamma}^{c}=0,\label{71}\\
v^{(2)M}_{6}\mathcal{Z}_{M}^{(2)}&=&-\partial_{i}\Gamma^{i}+\frac{l^{a}_{0}}{2\beta_{13}^{+}}\Phi_{2a}-\frac{e^{a}_{0}}{2\beta_{24}^{+}}\Phi_{4a}+\left(\frac{\beta_{4}}{\beta_{24}^{+}}-\frac{\beta_{1}}{\beta_{13}^{+}}\right)\varepsilon^{\alpha\beta\gamma}\epsilon_{abc}w_{\alpha}^{a}e_{\beta}^{b}l_{\gamma}^{c}\nonumber\\
&&+\left(\frac{\beta_{2}}{\beta_{24}^{+}}-\frac{\beta_{3}}{\beta_{13}^{+}}\right)\varepsilon^{\alpha\beta\gamma}\epsilon_{abc}A_{\alpha}^{a}e_{\beta}^{b}l_{\gamma}^{c}-\frac{3}{2}\varepsilon^{\alpha\beta\gamma}\epsilon_{abc}\left(\frac{\gamma_{1}}{\beta_{13}^{+}}e_{\alpha}^{a}-\frac{\gamma_{2}}{\beta_{24}^{+}}l_{\alpha}^{a}\right)e_{\beta}^{b}l_{\gamma}^{c}=0,\label{72}\\
v^{(2)M}_{7}\mathcal{Z}_{M}^{(2)}&=&\partial_{i}\Psi^{i}+\frac{1}{2}\frac{1}{\beta_{13}^{+}}\left(w^{a}_{0}-A^{a}_{0}\right)\Phi_{2a}+e^{a}_{0}\left(\frac{1}{\alpha_{2}}\Phi_{3a}-\frac{1}{\alpha_{1}}\Phi_{1a}\right)\nonumber\\
&&+\frac{\beta_{3}-\beta_{1}}{2\beta_{13}^{+}}\varepsilon^{\alpha\beta\gamma}\epsilon_{abc}\left(w_{\alpha}^{a}-A_{\alpha}^{a}\right)e_{\beta}^{b}\left(w_{\gamma}^{c}-A_{\gamma}^{c}\right)-\frac{3}{2}\frac{\gamma_{1}}{\beta_{13}^{+}}\varepsilon^{\alpha\beta\gamma}\epsilon_{abc}\left(w_{\alpha}^{a}-A_{\alpha}^{a}\right)e_{\beta}^{b}e_{\gamma}^{c}\nonumber\\
&&+\left(\frac{\beta_{1}}{\alpha_{1}}-\frac{\beta_{3}}{\alpha_{2}}\right)\varepsilon^{\alpha\beta\gamma}\epsilon_{abc}e_{\alpha}^{a}e_{\beta}^{b}e_{\gamma}^{c}+\left(\frac{\beta_{4}}{\alpha_{1}}-\frac{\beta_{2}}{\alpha_{2}}\right)\varepsilon^{\alpha\beta\gamma}\epsilon_{abc}e_{\alpha}^{a}l_{\beta}^{b}l_{\gamma}^{c}=0.\label{73}
\end{eqnarray}
After evaluating on the constraint surface, and taking into account that the equations for the Lagrange multipliers are fulfilled on such a surface, we can finally obtain the following three scalar equations:
\begin{eqnarray}
\mathcal{A}&=&\frac{1}{2}\frac{\beta_{2}-\beta_{4}}{\beta^{+}_{24}}\varepsilon^{\alpha\beta\gamma}\epsilon_{abc}\left(w_{\alpha}^{a}-A_{\alpha}^{a}\right)l_{\beta}^{b}\left(w_{\gamma}^{c}-A_{\gamma}^{c}\right)-\frac{3}{2}\frac{\gamma_{2}}{\beta^{+}_{24}}\varepsilon^{\alpha\beta\gamma}\epsilon_{abc}\left(w_{\alpha}^{a}-A_{\alpha}^{a}\right)l_{\beta}^{b}l_{\gamma}^{c}\nonumber\\
&&+\left(\frac{\beta_{1}}{\alpha_{1}}-\frac{\beta_{3}}{\alpha_{2}}\right)\varepsilon^{\alpha\beta\gamma}\epsilon_{abc}l_{\alpha}^{a}e_{\beta}^{b}e_{\gamma}^{c}+\left(\frac{\beta_{4}}{\alpha_{1}}-\frac{\beta_{2}}{\alpha_{2}}\right)\varepsilon^{\alpha\beta\gamma}\epsilon_{abc}l_{\alpha}^{a}l_{\beta}^{b}l_{\gamma}^{c}=0,\label{Al1}\\
\mathcal{B}&=&\left(\frac{\beta_{4}\beta_{3}-\beta_{1}\beta_{2}}{\beta^{+}_{24}\beta^{+}_{13}}\right)\varepsilon^{\alpha\beta\gamma}\epsilon_{abc}\left(w_{\alpha}^{a}-A_{\alpha}^{a}\right)e_{\beta}^{b}l_{\gamma}^{c}-\frac{3}{2}\varepsilon^{\alpha\beta\gamma}\epsilon_{abc}\left(\frac{\gamma_{1}}{\beta^{+}_{13}}e_{\alpha}^{a}-\frac{\gamma_{2}}{\beta^{+}_{24}}l_{\alpha}^{a}\right)e_{\beta}^{b}l_{\gamma}^{c}=0,\label{Al2}\\
\mathcal{C}&=&\frac{1}{2}\frac{\beta_{3}-\beta_{1}}{\beta^{+}_{13}}\varepsilon^{\alpha\beta\gamma}\epsilon_{abc}\left(w_{\alpha}^{a}-A_{\alpha}^{a}\right)e_{\beta}^{b}\left(w_{\gamma}^{c}-A_{\gamma}^{c}\right)-\frac{3}{2}\frac{\gamma_{1}}{\beta^{+}_{13}}\varepsilon^{\alpha\beta\gamma}\epsilon_{abc}\left(w_{\alpha}^{a}-A_{\alpha}^{a}\right)e_{\beta}^{b}e_{\gamma}^{c}\nonumber\\
&&+\left(\frac{\beta_{1}}{\alpha_{1}}-\frac{\beta_{3}}{\alpha_{2}}\right)\varepsilon^{\alpha\beta\gamma}\epsilon_{abc}e_{\alpha}^{a}e_{\beta}^{b}e_{\gamma}^{c}+\left(\frac{\beta_{4}}{\alpha_{1}}-\frac{\beta_{2}}{\alpha_{2}}\right)\varepsilon^{\alpha\beta\gamma}\epsilon_{abc}e_{\alpha}^{a}l_{\beta}^{b}l_{\gamma}^{c}=0.\label{Al3}
\end{eqnarray}
These scalar equations mix Lagrange multipliers with dynamical variables so that they do not lead to new constraints. Hence, our procedure for finding a full set of constraints for general exotic bi-gravity has finished.  Note that in the limit that $\beta_{3}\rightarrow0$ or $\beta_{4}\rightarrow0$, the  equations (\ref{71})-(\ref{73}) reduce to two scalar equation:

\textit{(i)} $\beta_{3}=0$. Here $\Theta^{0}$ and  $\Gamma^{0}$ are secondary constraints while $\Theta^{i}=0$ and $\Gamma^{i}=0$ are equations for Lagrange multipliers, so the corresponding scalar expressions  hence take the form
\begin{eqnarray}
\mathcal{A}&=&\frac{1}{2}\frac{\beta_{2}-\beta_{4}}{\beta^{+}_{24}}\varepsilon^{\alpha\beta\gamma}\epsilon_{abc}\left(w_{\alpha}^{a}-A_{\alpha}^{a}\right)l_{\beta}^{b}\left(w_{\gamma}^{c}-A_{\gamma}^{c}\right)-\frac{3}{2}\frac{\gamma_{2}}{\beta^{+}_{24}}\varepsilon^{\alpha\beta\gamma}\epsilon_{abc}\left(w_{\alpha}^{a}-A_{\alpha}^{a}\right)l_{\beta}^{b}l_{\gamma}^{c}\nonumber\\
&&+\frac{\beta_{1}}{\alpha_{1}}\varepsilon^{\alpha\beta\gamma}\epsilon_{abc}l_{\alpha}^{a}e_{\beta}^{b}e_{\gamma}^{c}+\left(\frac{\beta_{4}}{\alpha_{1}}-\frac{\beta_{2}}{\alpha_{2}}\right)\varepsilon^{\alpha\beta\gamma}\epsilon_{abc}l_{\alpha}^{a}l_{\beta}^{b}l_{\gamma}^{c}=0,\\
\mathcal{B}&=&-\frac{\beta_{2}}{\beta^{+}_{24}}\varepsilon^{\alpha\beta\gamma}\epsilon_{abc}\left(w_{\alpha}^{a}-A_{\alpha}^{a}\right)e_{\beta}^{b}l_{\gamma}^{c}-\frac{3}{2}\varepsilon^{\alpha\beta\gamma}\epsilon_{abc}\left(\frac{\gamma_{1}}{\beta_{1}}e_{\alpha}^{a}-\frac{\gamma_{2}}{\beta^{+}_{24}}l_{\alpha}^{a}\right)e_{\beta}^{b}l_{\gamma}^{c}=0.
\end{eqnarray}

\textit{(ii)} $\beta_{4}=0$. Here,  $\Psi^{0}$ and $\Gamma^{0}$ are secondary constraints while $\Psi^{i}=0$ and $\Gamma^{i}=0$ are equations for Lagrange multipliers. Then the scalar expressions  are given by
\begin{eqnarray}
\mathcal{B}&=&-\frac{\beta_{1}}{\beta^{+}_{13}}\varepsilon^{\alpha\beta\gamma}\epsilon_{abc}\left(w_{\alpha}^{a}-A_{\alpha}^{a}\right)e_{\beta}^{b}l_{\gamma}^{c}-\frac{3}{2}\varepsilon^{\alpha\beta\gamma}\epsilon_{abc}\left(\frac{\gamma_{1}}{\beta^{+}_{13}}e_{\alpha}^{a}-\frac{\gamma_{2}}{\beta_{2}}l_{\alpha}^{a}\right)e_{\beta}^{b}l_{\gamma}^{c}=0,\\
\mathcal{C}&=&\frac{1}{2}\frac{\beta_{3}-\beta_{1}}{\beta^{+}_{13}}\varepsilon^{\alpha\beta\gamma}\epsilon_{abc}\left(w_{\alpha}^{a}-A_{\alpha}^{a}\right)e_{\beta}^{b}\left(w_{\gamma}^{c}-A_{\gamma}^{c}\right)-\frac{3}{2}\frac{\gamma_{1}}{\beta^{+}_{13}}\varepsilon^{\alpha\beta\gamma}\epsilon_{abc}\left(w_{\alpha}^{a}-A_{\alpha}^{a}\right)e_{\beta}^{b}e_{\gamma}^{c}\nonumber\\
&&+\left(\frac{\beta_{1}}{\alpha_{1}}-\frac{\beta_{3}}{\alpha_{2}}\right)\varepsilon^{\alpha\beta\gamma}\epsilon_{abc}e_{\alpha}^{a}e_{\beta}^{b}e_{\gamma}^{c}-\frac{\beta_{2}}{\alpha_{2}}\varepsilon^{\alpha\beta\gamma}\epsilon_{abc}e_{\alpha}^{a}l_{\beta}^{b}l_{\gamma}^{c}=0.
\end{eqnarray}

Note that the secondary constraints, in each case, and the primary ones define the physical phase space $\Gamma_{\text{Phys}}$, which is a submanifold of the initial phase space $\Gamma$. In this setting, the time evolution of the physical degrees of freedom takes place entirely in $\Gamma_{\text{Phys}}$. On the other hand, although the above derived scalar equations are not new constraints, they should give information about the dynamical content of general exotic bi-gravity. From this perspective, we could use these scalar relations to study the AdS background.

\subsection{Application for the Anti de Sitter background}
Let us analyze the scalar equations given in (\ref{Al1})-(\ref{Al3}) around a common maximally symmetric background with a cosmological constant. Following Ref. \cite{GEMG}, we suppose that $\bar{e}$ and $\bar{w}$ are the dreibein and the dualized spin-connection for the Anti-de Sitter background. Then we can take the following ansatz:
 \begin{equation}
e_{a}^{\mu}=a_{1}\bar{e}_{a}^{\mu},\quad w_{a}^{\mu}=\bar{w}+b_{1}\bar{e}_{a}^{\mu},\quad
l_{a}^{\mu}=a_{2}\bar{e}_{a}^{\mu},\quad A_{a}^{\mu}=\bar{w}+b_{2}\bar{e}_{a}^{\mu},\label{anzatz}
\end{equation}
where  $a_{1}$, $a_{2}$, $b_{1}$, and $b_{2}$, are constant parameters. Now the aim is to determine these constants in terms of the parameters of the theory. Plugging the ansatz (\ref{anzatz}) into (\ref{Al1}) or (\ref{Al3}) leads to the following expression,  
\begin{equation}
a_{1}^{2}\left(\frac{9}{8}\frac{\left(\beta_{2}^{2}-\beta_{4}^{2}\right)}{\beta^{2}}\gamma_{1}^{2}+\frac{\beta_{1}}{\alpha_{1}}-\frac{\beta_{3}}{\alpha_{2}}\right)
+a_{2}^{2}\left(\frac{9}{8}\frac{\left(\beta_{3}^{2}-\beta_{1}^{2}\right)}{\beta^{2}}\gamma_{2}^{2}+\frac{\beta_{4}}{\alpha_{1}}-\frac{\beta_{2}}{\alpha_{2}}\right)
+\frac{9}{4}\frac{\left(\beta_{1}\beta_{4}-\beta_{2}\beta_{3}\right)}{\beta^{2}}\gamma_{1}\gamma_{2}a_{1}a_{2}=0,\label{anzatz2}
\end{equation}
where $\beta=\beta_{4}\beta_{3}-\beta_{1}\beta_{2}$. Whereas, by substituting Eq. (\ref{anzatz}) into Eq. (\ref{Al2}) we obtain:
\begin{equation}
\beta_{1}\left(3\gamma_{2}a_{2}+2\beta_{2}b_{2}\right)-\beta_{2}\left(3\gamma_{1}a_{1}+2\beta_{1}b_{1}\right)+\beta_{3}\left(3\gamma_{2}a_{2}+2\beta_{4}b_{1}\right)-\beta_{4}\left(3\gamma_{1}a_{1}+2\beta_{3}b_{2}\right)=0.\label{anzatz3}
\end{equation}
It is obvious that for the generic case, where all the coupling constants $\beta$'s are non-vanishing, the last equation (\ref{anzatz3}) is satisfied only if
\begin{eqnarray}
3\gamma_{2}a_{2}+2\beta_{2}b_{2}&=&0,\label{gamm1}\\
3\gamma_{2}a_{2}+2\beta_{4}b_{1}&=&0,\label{gamm2}\\
3\gamma_{1}a_{1}+2\beta_{1}b_{1}&=&0,\label{gamm3}\\
3\gamma_{1}a_{1}+2\beta_{3}b_{2}&=&0.\label{gamm4}
\end{eqnarray}
So, by combining (\ref{gamm1}) with (\ref{gamm2}), and  (\ref{gamm3}) with (\ref{gamm4}), we  find 
\begin{equation}
\gamma_{1}=-\frac{1}{3}\frac{\left(\beta_{1}b_{1}+\beta_{3}b_{2}\right)}{a_{1}},\quad
\gamma_{2}=-\frac{1}{3}\frac{\left(\beta_{2}b_{2}+\beta_{4}b_{1}\right)}{a_{2}}\label{gamas}
\end{equation}
It is worth noticing that, $\gamma_{1}$ and $\gamma_{2}$ for the cases \textit{(i)} and \textit{(ii)} can be obtained from these expressions upon setting the limit $\beta_{3}\rightarrow0$ and $\beta_{4}\rightarrow0$, respectively:

\textit{(i)} If $\beta_{3}\rightarrow0$, then,
\begin{equation}
\gamma_{1}=-\frac{1}{3}\frac{\beta_{1}b_{2}}{a_{1}},\quad
\gamma_{2}=-\frac{1}{3}\frac{\left(\beta_{2}b_{2}+\beta_{4}b_{1}\right)}{a_{2}}.
\end{equation}
Thus, up to a $2$ factor, these results coincide with those reported in \cite{GEMG}. Similarly, if we choose $\beta_{1}\rightarrow0$, then we obtain
\begin{equation}
\gamma_{1}=-\frac{1}{3}\frac{\beta_{3}b_{1}}{a_{1}},\quad
\gamma_{2}=-\frac{1}{3}\frac{\left(\beta_{2}b_{2}+\beta_{4}b_{1}\right)}{a_{2}}.
\end{equation}

\textit{(ii)} If $\beta_{4}\rightarrow0$, then,
\begin{equation}
\gamma_{1}=-\frac{1}{3}\frac{\left(\beta_{1}b_{1}+\beta_{3}b_{2}\right)}{a_{1}},\quad
\gamma_{2}=-\frac{1}{3}\frac{\beta_{2}b_{2}}{a_{2}}.
\end{equation}
Similarly, if we choose $\beta_{2}=0$, then we obtain
\begin{equation}
\gamma_{1}=-\frac{1}{3}\frac{\left(\beta_{1}b_{1}+\beta_{3}b_{2}\right)}{a_{1}},\quad
\gamma_{2}=-\frac{1}{3}\frac{\beta_{4}b_{1}}{a_{2}}.
\end{equation}

Focusing on the general expressions for $\gamma_{1}$ and $\gamma_{2}$, given in (\ref{gamas}), then substituting them into Eq. (\ref{anzatz2}), we obtain the following equation:
\begin{equation}
\left[\frac{1}{8}b_{1}^{2}+a_{1}^{2}\left(\frac{\beta_{1}}{\alpha_{1}}-\frac{\beta_{3}}{\alpha_{2}}\right)\right]-\left[\frac{1}{8}b_{2}^{2}+a_{2}^{2}\left(\frac{\beta_{2}}{\alpha_{2}}-\frac{\beta_{4}}{\alpha_{1}}\right)\right]=0.
\end{equation}
Multiplying this equation by $\alpha_{1}\alpha_{2}$ the resulting equation now leads to
\begin{equation}
\alpha_{2}\left(\frac{1}{8}\alpha_{1}b_{1}^{2}+\beta_{1}a_{1}^{2}+\beta_{4}a_{2}^{2}\right)-\alpha_{1}\left(\frac{1}{8}\alpha_{2}b_{2}^{2}+\beta_{2}a_{2}^{2}+\beta_{3}a_{1}^{2}\right)=0.\label{alfas}
\end{equation}
Note that this last equation requires,
\begin{eqnarray}
\frac{1}{8}\alpha_{1}b_{1}^{2}+\beta_{1}a_{1}^{2}+\beta_{4}a_{2}^{2}&=&-\frac{\Lambda}{8}\alpha_{1},\\
\frac{1}{8}\alpha_{2}b_{2}^{2}+\beta_{2}a_{2}^{2}+\beta_{3}a_{1}^{2}&=&-\frac{\Lambda}{8}\alpha_{2},
\end{eqnarray}
where $\Lambda$ is a constant, which can be identified as the cosmological constant for the Anti-de Sitter background; $\Lambda= -1/\ell^{2}$ where $\ell$ is the Anti-de Sitter space radius. With this, we immediately find
\begin{equation}
\alpha_{1}=-8\frac{\beta_{1}a_{1}^{2}+\beta_{4}a_{2}^{2}}{\Lambda+b_{1}^{2}},\quad
\alpha_{2}=-8\frac{\beta_{2}a_{2}^{2}+\beta_{3}a_{1}^{2}}{\Lambda+b_{2}^{2}}.\label{alfasfinales}
\end{equation}
Finally, from these generic expressions (\ref{alfasfinales}), we can easily derive equations for cases \textit{(i)} and \textit{(ii)}, respectively:

\textit{ (i)} If $\beta_{3}\rightarrow0$, then,
\begin{equation}
\alpha_{1}=-8\frac{\beta_{1}a_{1}^{2}+\beta_{4}a_{2}^{2}}{\Lambda+b_{1}^{2}},\quad
\alpha_{2}=-8\frac{\beta_{2}a_{2}^{2}}{\Lambda+b_{2}^{2}}.
\end{equation}
Thus, up to a $1/4$ factor, these results coincide with those reported in Ref. \cite{GEMG}. Back in (\ref{alfasfinales}) we obtain the corresponding equations when $\beta_{1}\rightarrow0$,
\begin{equation}
\alpha_{1}=-8\frac{\beta_{4}a_{2}^{2}}{\Lambda+b_{1}^{2}},\quad
\alpha_{2}=-8\frac{\beta_{2}a_{2}^{2}+\beta_{3}a_{1}^{2}}{\Lambda+b_{2}^{2}}.
\end{equation}

\textit{ (ii)} If $\beta_{4}\rightarrow0$, then,
\begin{equation}
\alpha_{1}=-8\frac{\beta_{1}a_{1}^{2}}{\left(\Lambda+b_{1}^{2}\right)},\quad
\alpha_{2}=-8\frac{\beta_{2}a_{2}^{2}+\beta_{3}a_{1}^{2}}{\left(\Lambda+b_{2}^{2}\right)}.
\end{equation}
Back in (\ref{alfasfinales}), after by setting $\beta_{2}=0$, we also obtain
\begin{equation}
\alpha_{1}=-8\frac{\beta_{1}a_{1}^{2}+\beta_{4}a_{2}^{2}}{\Lambda+b_{1}^{2}},\quad
\alpha_{2}=-8\frac{\beta_{3}a_{1}^{2}}{\Lambda+b_{2}^{2}}.\label{}
\end{equation}

\section{Gauge symmetry and degrees of freedom}

\subsection{Gauge transformations and their generators}
\label{Symmetries}
After completing the constraints analysis, we now wish to compute the generator of the gauge symmetry `$\delta_{G}$'. Usually, after finding the complete set of constraints of the theory, what remains is to introduce them into the initial action through a set of arbitrary Lagrange multipliers \cite{Henneaux,Rothe}. Then, all the constraints will be enforced by the dynamics, i.e., the equations of motion. Notice, however, that because of the dynamical character of the constraints, we also need to require their stability along the time evolution of the system. Taking into account the consistency over time of the constraints and  that a Lagrangian is defined up to total time derivatives, it follows that a term such as $d(\lambda^{M}\overline{\Xi}_{M})/dt=\dot{\lambda}^{M}\overline{\Xi}_{M}+\lambda^{M}\dot{\overline{\Xi}}_{M}$ does not affect the dynamics of the system, so the  action (\ref{action}) can be written as
\begin{equation}
\tilde{S}[\xi,\lambda]=\int\left(a^{I}[\xi]\dot{\xi}_{I}-\dot{\lambda}^{M}\overline{\Xi}_{M}[\xi]-\mathcal{H}[\xi]\right)dtd^{2}x.\label{actionfinal}
\end{equation}
Here the set of Lagrange multipliers and of constraints are given by
\begin{eqnarray}
\lambda^{M}&=&\left(\lambda^{1}_{a},\lambda^{2}_{a},\lambda^{3}_{a},\lambda^{4}_{a},\lambda^{5},\lambda^{6},\lambda^{7}\right),\\
\overline{\Xi}_{M}&=&\left(\Phi_{1}^{a},\Phi_{2}^{a},\Phi_{3}^{a},\Phi_{4}^{a},\Psi^{0},\Theta^{0},\Gamma^{0}\right),
\end{eqnarray}
while $a^{I}$ and $\xi_{I}$ are given by (\ref{var}) and (\ref{1form}), respectively.  In that way, each one of the consistency conditions for the constraints present in the theory will directly become an equation of motion. Relatedly, we can observe that upon the redefinition of the Lagrange multipliers, $\dot{\lambda}^{M}\rightarrow-\eta^{M}$ \cite{Barcelos1, Barcelos2}, the action (\ref{actionfinal}) takes the form $\tilde{S}=\int\left(a^{I}\dot{\xi}_{I}+\eta^{M}\overline{\Xi}_{M}-\mathcal{H}\right)dtd^{2}x$. As a result, the variation of this action with respect to $\eta^{M}$ will give rise to all the constraints $\overline{\Xi}_{M}=0$.
Moreover, it is straightforward to show that the canonical Hamiltonian $\mathcal{H}$, as one would expect, will drop from the action after being evaluated on the constraint surface $\Gamma_{\text{Phys}}$,
\begin{equation}
\mathcal{H}=\left(w_{a0}\Phi^{a}_{1}+e_{a0}\Phi^{a}_{2}+A_{a0}\Phi^{a}_{3}+l_{a0}\Phi^{a}_{4}\right)|_{\Gamma_{\text{Phys}}}=0.
\end{equation}
This fact shows the general covariance of the theory, and therefore, the dynamics will be governed by the constraints. At this stage, the new action (\ref{actionfinal}) explicitly contains the necessary information to describe all the dynamics of the theory that now take place on the physical phase space $\Gamma_{\text{Phys}}$.

It is worth noticing that the variables $w^{a}_{0}$, $e^{a}_{0}$, $A^{a}_{0}$, and $l^{a}_{0}$ already represent a set of Lagrange multipliers implementing the constraints $\Phi^{a}_{1}$, $\Phi^{a}_{2}$, $\Phi^{a}_{3}$, and $\Phi^{a}_{4}$, respectively, meaning that there is no need to introduce new Lagrange multipliers for all constraints found. In fact, by redefining $w^{a}_{0}\rightarrow\dot{\lambda}^{a}_{1}$, $e^{a}_{0}\rightarrow\dot{\lambda}^{a}_{2}$, $A^{a}_{0}\rightarrow\dot{\lambda}^{a}_{3}$, and $l^{a}_{0}\rightarrow\dot{\lambda}^{a}_{4}$, we just need to introduce three new Lagrange multipliers associated with $\Psi^{0}$, $\Theta^{0}$, and $\Gamma^{0}$, and we  name them $\dot{\lambda}^{5}$, $\dot{\lambda}^{6}$, and $\dot{\lambda}^{7}$.  In that way, we find that the dynamics is now  governed by the following action principle:
\begin{eqnarray}
\widetilde{S}&=&\int\Bigl[\varepsilon^{0ij}\left(\frac{1}{2}\alpha_{1}\dot{w}_{ai}w^{a}_{j}+\beta_{13}^{+}\dot{e}_{ai}e^{a}_{j}+\frac{1}{2}\alpha_{2}\dot{A}_{ai}A^{a}_{j}+\beta_{24}^{+}\dot{l}_{ai}l^{a}_{j}\right)- \dot{\lambda}^{1}_{a}\Phi^{a}_{1}-\dot{\lambda}^{2}_{a}\Phi^{a}_{2}-\dot{\lambda}^{3}_{a}\Phi^{a}_{3}-\dot{\lambda}^{4}_{a}\Phi^{a}_{4}\Bigr.\nonumber\\
&&\Bigl.-\dot{\lambda}^{5}\Psi^{0}-\dot{\lambda}^{6}\Theta^{0}-\dot{\lambda}^{7}\Gamma^{0}\Bigr]dtd^{2}x,\label{new}
\end{eqnarray}
Now, we can take the canonical variable set:
\begin{equation}
\widetilde{\xi}_{I}=\left(\overline{\xi}_{I},\lambda_{M}\right),\label{107}
\end{equation}
with $\overline{\xi}_{I}=\left(w_{ai},e_{ai},A_{ai},l_{ai}\right)$. Then the corresponding components of one-form are:
\begin{equation}
\widetilde{a}^{I}=\left(\overline{a}^{I},-\overline{\Xi}_{M}\right),\label{108}
\end{equation}
with $\overline{a}^{I}=\varepsilon^{0ij}\left(\frac{1}{2}\alpha_{1}w_{j}^{a},\beta^{+}_{13}e_{j}^{a},\frac{1}{2}\alpha_{2}A_{j}^{a},\beta^{+}_{24}l_{j}^{a}\right)$. Making use of the definition of the symplectic matrix in Eq. (\ref{F}), and after a straightforward computation, we can see that in terms of $\overline{\xi}_{I}$ and $\overline{a}^{I}$, the resulting square matrix takes the block form,
 \begin{equation}
\widetilde{\mathcal{F}}_{KJ}[\overline{\xi}]=
\left(
\begin{array}{ccc} 
\left(\frac{\delta \overline{a}_{J}[x] }{\delta\overline{\xi}^{I}[y]}-\frac{\delta \overline{a}_{I}[y]}{\delta\overline{\xi}^{J}[x]}\right)&- \left(\frac{\delta\overline{\Xi}_{M}[x]}{\delta\overline{\xi}^{J}[y]}\right)^{T}\\ 
\left(\frac{\delta\overline{\Xi}_{M}[y]}{\delta\overline{\xi}^{J}[x]}\right)&0\\
\end{array}
\right),\label{finalmatrix}
\end{equation}
where
\begin{equation}
\left(\frac{\delta \overline{a}_{J}[x] }{\delta\overline{\xi}^{I}[y]}-\frac{\delta \overline{a}_{I}[y]}{\delta\overline{\xi}^{J}[x]}\right)=-\varepsilon^{0ij}
\left(
  \begin{array}{cccccccc}
  \alpha_{1}\eta^{ab}   &   0    & 0 &   0\\                                                                   
  0   &    2\beta^{+}_{13}\eta^{ab}    &     0&    0 	\\
 0  &  0  &  \alpha_{2}\eta^{ab} 	 &    0\\
  0  &  0   &  0 	&   2\beta^{+}_{24}\eta^{ab}
  \end{array}
\right)\delta^{2}(x-y),
\end{equation}
and
\begin{equation}
\left(\frac{\delta\overline{\Xi}_{M}[y]}{\delta\overline{\xi}^{J}[x]}\right)=-2\varepsilon^{0ij}
\left(
  \begin{array}{cccccccc}
   \frac{\alpha_{1}}{2}\mathbf{W}^{\textbf{x}ab}_{i} &  \beta_{1}\mathbf{E}_{i}^{ab} &0 &\beta_{4}\mathbf{L}_{i}^{ab}  \\
 \beta_{1}\mathbf{E}_{i}^{ab} & \beta_{1}\mathbf{W}_{i}^{\textbf{x}ab}+\beta_{3}\mathbf{A}_{i}^{\textbf{x}ab}+3\gamma_{1}\mathbf{E}_{i}^{ab} &  \beta_{3}\mathbf{E}_{i}^{ab} &0\\
0&\beta_{3}\mathbf{E}_{i}^{ab}&\frac{\alpha_{2}}{2}\mathbf{A}_{i}^{\textbf{x}ab}&\beta_{2}\mathbf{L}_{i}^{ab}\\
\beta_{4}\mathbf{L}_{i}^{ab}&0&\beta_{2}\mathbf{L}_{i}^{ab}&\beta_{2}\mathbf{A}_{i}^{\textbf{x}ab}+\beta_{4}\mathbf{W}_{i}^{\textbf{x}ab}+3\gamma_{2}\mathbf{L}_{i}^{ab}\\
-\frac{1}{2}l^{b}_{i}&0&\frac{1}{2}l_{i}^{b}&\frac{1}{2}\left(w_{i}^{b}-A_{i}^{b}\right)\\
0&\frac{1}{2}l_{i}^{b}&0&-\frac{1}{2}e_{i}^{b}\\
-\frac{1}{2}e_{i}^{b}&\frac{1}{2}\left(w_{i}^{b}-A_{i}^{b}\right)&\frac{1}{2}e_{i}^{b}&0
 \end{array}
\right)\delta^{2}(x-y).\label{FInalMatrix}
\end{equation}

Again, this matrix is a singular one and  has zero-modes with the following structure:
\begin{eqnarray}
\widetilde{v}^{I}_{(1)}&=&\Bigl(-\mathbf{W}^{\textbf{x}ab}_{i}, -\frac{\beta_{1}}{\beta^{+}_{13}}\mathbf{E}^{ab}_{i},0,-\frac{\beta_{4}}{\beta^{+}_{24}}\mathbf{L}^{ab}_{i},\eta^{ab},0,0,0,0,0,0\Bigr)\delta^{2}(x-y),\\
\widetilde{v}^{I}_{(2)}&=&\Bigl(-2\frac{\beta_{1}}{\alpha_{1}}\mathbf{E}^{ab}_{i},-\frac{1}{\beta_{13}}\left(\beta_{1}{\mathbf{W}}^{\textbf{x}ab}_{i}+\beta_{3}{\mathbf{A}}^{\textbf{x}ab}_{i}+3\gamma_{1}{\mathbf{E}}^{ab}_{i}\right),-2\frac{\beta_{3}}{\alpha_{2}}\mathbf{E}^{ab}_{i},0,0,\eta^{ab},0,0,0,0,0\Bigr)\delta^{2}(x-y)\\
\widetilde{v}^{I}_{(3)}&=&\Bigl(0, -\frac{\beta_{3}}{\beta^{+}_{13}}\mathbf{E}^{ab}_{i},-\mathbf{A}^{\textbf{x}ab}_{i},-\frac{\beta_{2}}{\beta^{+}_{24}}\mathbf{L}^{ab}_{i},0,0,\eta^{ab},0,0,0,0\Bigr)\delta^{2}(x-y),\\
\widetilde{v}^{I}_{(4)}&=&\Bigl(-2\frac{\beta_{4}}{\alpha_{1}}\mathbf{L}^{ab}_{i},0,-2\frac{\beta_{2}}{\alpha_{2}}\mathbf{L}^{ab}_{i},-\frac{1}{\beta_{24}}\left(\beta_{4}{\mathbf{W}}^{\textbf{x}ab}_{i}+\beta_{2}{\mathbf{A}}^{\textbf{x}ab}_{i}+3\gamma_{2}{\mathbf{L}}^{ab}_{i}\right), 0,0,0,\eta^{ab},0,0,0\Bigr)\delta^{2}(x-y),\\
\widetilde{v}^{I}_{(5)}&=&\Bigl(\frac{1}{\alpha_{1}}l^{a}_{i},0,-\frac{1}{\alpha_{2}}l^{a}_{i} ,-\frac{1}{2\beta_{24}}\left(w^{a}_{i}-A^{a}_{i}\right), 0,0,0,0,1,0,0\Bigr)\delta^{2}(x-y),\\
\widetilde{v}^{I}_{(6)}&=&\Bigl(0,-\frac{1}{2\beta_{13}}l^{a}_{i},0,\frac{1}{2\beta_{24}}e^{a}_{i},0, 0,0,0,0,1,0\Bigr)\delta^{2}(x-y),\\
\widetilde{v}^{I}_{(7)}&=&\Bigl(\frac{1}{\alpha_{1}}e^{a}_{i},-\frac{1}{2\beta_{13}}\left(w^{a}_{i}-A^{a}_{i}\right),-\frac{1}{\alpha_{2}}e^{a}_{i},0 ,0, 0,0,0,0,0,1\Bigr)\delta^{2}(x-y).
\end{eqnarray}
With these results in hand, we find that, on $\Gamma_{\text{Phys}}$, only the zero-modes, $\widetilde{v}^{I}_{(1)}$, $\widetilde{v}^{I}_{(2)}$, $\widetilde{v}^{I}_{(3)}$, and $\widetilde{v}^{I}_{(4)}$, are orthogonal to the gradient of $\mathcal{H}$, i.e., $\int \widetilde{v}^{I}_{(n)}d^{2}y\int \left(\delta\mathcal{H}/\delta\widetilde{\xi}^{I}\right)d^{2}x\, |_{\Gamma_{\text{Phys}}}=0$. In turn, this implies that such zero-modes produce  isopotential displacements on $\mathcal{H}$. And as has been pointed out in Refs. \cite{Omar,Montani1,Montani2},  this happens because of an existing symmetry in the canonical Hamiltonian $\mathcal{H}$, a situation typical of gauge  invariant theories. As a result, these zero-modes must be the  canonical generators of the gauge symmetries of the canonical variables $\widetilde{\xi}^{I}$ on the physical phase space $\Gamma_{\text{Phys}}$.

According to Refs. \cite{Montani1,Montani2}, all of the local infinitesimal transformations of the canonical variables $\widetilde{\xi}_{I}$  should be found through the following gauge generating functional:
\begin{equation}
\delta_{G}\widetilde{\xi}^{I}[y]=\int \widetilde{v}_{(n)}[x,y]\eta^{(n)}[x]d^{2}x.\label{Generator}
\end{equation}
Here, $\widetilde{v}_{(n)}$ are the zero-modes generating local displacements on the isopotential surface and $\eta^{(n)}$ is a set of arbitrary infinitesimal parameters, one for each of the zero-modes. In this way, one can infer from Eq. (\ref{Generator}), that the infinitesimal gauge transformations of our dynamical variables take the form
\begin{eqnarray}
\delta_{G}w^{a}_{i}&=&D_{i}\eta^{(1)\, a}-\frac{2}{\alpha_{1}}\epsilon^{abc}\left[\beta_{1}e_{ci}\eta^{(2)}_{b}+\beta_{4}l_{ci}\eta^{(4)}_{b}\right],\label{gauge1}\\
\delta_{G}e^{a}_{i}&=&D_{i}\eta^{(2)\, a}+\frac{1}{\beta^{+}_{13}}\epsilon^{abc}\left[\beta_{3}\left(A_{bi}-w_{bi}\right)\eta^{(2)}_{c}-3\gamma_{1}e_{bi}\eta^{(2)}_{c}+\beta_{1}e_{bi}\eta^{(1)}_{c}-\beta_{3}e_{bi}\eta^{(3)}_{c}\right],\label{gauge2}\\
\delta_{G}A^{a}_{i}&=&\nabla_{i}\eta^{(3)\, a}-\frac{2}{\alpha_{2}}\epsilon^{abc}\left[\beta_{3}e_{ci}\eta^{(2)}_{b}+\beta_{2}l_{ci}\eta^{(4)}_{b}\right],\label{gauge3}\\
\delta_{G}l^{a}_{i}&=&\nabla_{i}\eta^{(4)\, a}-\frac{1}{\beta^{+}_{24}}\epsilon^{abc}\left[\beta_{4}\left(A_{bi}-w_{bi}\right)\eta^{(4)}_{c}+3\gamma_{2}l_{bi}\eta^{(4)}_{c}+\beta_{4}l_{bi}\eta^{(1)}_{c}+\beta_{2}l_{bi}\eta^{(3)}_{c}\right].\label{gauge4}
\end{eqnarray}
Whereas the gauge transformations of the Lagrange multipliers $\lambda_{1}^{a}$, $\lambda_{2}^{a}$, $\lambda_{3}^{a}$, and $\lambda_{4}^{a}$ turn out to be
\begin{eqnarray}
\delta_{G}\lambda_{1}^{a}&=&\eta^{(1)a},\\
\delta_{G}\lambda_{2}^{a}&=&\eta^{(2)a},\\
\delta_{G}\lambda_{3}^{a}&=&\eta^{(3)a},\\
\delta_{G}\lambda_{4}^{a}&=&\eta^{(4)a}.
\end{eqnarray}
By turning back to the original variables, $\dot{\lambda}^{a}_{1}\rightarrow w^{a}_{0}$, $\dot{\lambda}^{a}_{2}\rightarrow e^{a}_{0}$, $\dot{\lambda}^{a}_{3}\rightarrow A^{a}_{0}$, and $\dot{\lambda}^{a}_{4}\rightarrow l^{a}_{0}$, we find the gauge transformations for the initial set of Lagrange multipliers
\begin{eqnarray}
\delta_{G}w_{0}^{a}&=&\dot{\eta}^{(1)a},\\
\delta_{G}e_{0}^{a}&=&\dot{\eta}^{(2)a},\\
\delta_{G}A_{0}^{a}&=&\dot{\eta}^{(3)a},\\
\delta_{G}l_{0}^{a}&=&\dot{\eta}^{(4)a}.
\end{eqnarray}
Meanwhile, the remaining gauge variations for $\lambda^{5}$, $\lambda^{6}$, and $\lambda^{7}$, are equal to zero. Note that `$\delta_{G}$' corresponds to the fundamental gauge symmetries of the theory, but not to diffeomorphisms `$\delta_{\text{diff}}$'. This is obtained in the next subsection.

\subsection{Diffeomorphism symmetry}
\label{Diffeo}
Now, it is well-known that an appropriate choice of gauge parameters does generate the diffeomorphism symmetries. Let us redefine the gauge parameters:
\begin{eqnarray}
\eta^{(1)\, a}=w^{a}_{i}\zeta^{i},&&\quad\eta^{(2)\, a}=e^{a}_{j}\zeta^{j},\label{gaugeparameter1}\\
\quad\eta^{(3)\, a}=A^{a}_{i}\zeta^{i},&&\quad\eta^{(4)\, a}=l^{a}_{i}\zeta^{i},\label{gaugeparameter2}
\end{eqnarray}
where $\zeta^{i}$ is  an arbitrary two-vector. Substitution of  (\ref{gaugeparameter1}) and (\ref{gaugeparameter2}) into the gauge transformations Eqs. (\ref{gauge1})-(\ref{gauge4}) yields  spatial diffeomorphisms symmetry for the dynamical variables,
\begin{eqnarray}
\delta_{\text{diff}}w^{a}_{i}&=&\pounds_{\zeta}w^{a}_{i}+\frac{1}{\alpha_{1}}\varepsilon_{0ij}\zeta^{j}\Phi^{a}_{1},\\
\delta_{\text{diff}}e^{a}_{i}&=&\pounds_{\zeta}e^{a}_{i}+\frac{1}{\beta^{+}_{13}}\varepsilon_{0ij}\zeta^{j}\Phi^{a}_{2},\\
\delta_{\text{diff}}A^{a}_{i}&=&\pounds_{\zeta}A^{a}_{i}+\frac{1}{\alpha_{2}}\varepsilon_{0ij}\zeta^{j}\Phi^{a}_{3},\\
\delta_{\text{diff}}l^{a}_{i}&=&\pounds_{\zeta}l^{a}_{i}+\frac{1}{\beta^{+}_{24}}\varepsilon_{0ij}\zeta^{j}\Phi^{a}_{4},
\end{eqnarray}
up to a constraint function which vanishes on $\Gamma_{\text{Phys}}$. There, $\pounds_{\zeta}$ is the Lie derivative concerning the vector field $\zeta^{i}$. Thus, the diffeomorphism are not an independent symmetry; they are contained on the fundamental gauge transformations `$\delta_{G}$'.

Let us emphasize that we have identified the correct transformation rules on the dynamical variables of the theory, by using the zero-modes of the pre-symplectic matrix (\ref{finalmatrix}) only.

\subsection{Physical degrees of freedom}
\label{DoF}
We will use the results obtained in the preceding section  to perform the counting of physical degrees of freedom of the theory. With respect to the constraints, let us shortly remember that we obtained $12$ primary constraints: $\Phi^{a}_{1}$, $\Phi^{a}_{2}$, $\Phi^{a}_{3}$, $\Phi^{a}_{4}$. Then, by considering the zero-modes of $\mathcal{F}_{KJ}^{(1)}$ in Eq. (\ref{secondEM}), we found a set of $9$ integrability conditions: $\mathcal{X}^{d}$, $\mathcal{Y}^{d}$, $\mathcal{Z}^{d}$. From such as conditions, we considered two equivalent cases where there are 2 secondary costraints. For instance, if $l^{\mu}_{d}$ is invertible and $\beta_{3}$ is zero, then the theory has  $2$ secondary constraints: $\Theta^{0}$, $\Gamma^{0}$. On the other hand, after the (2+1) decomposition of  our fields, the $24$ spatial components of the dreibeins and the spin-connections $(e_{i}^{a}, w_{i}^{a},l_{i}^{a},A_{i}^{a})$ are dynamical variables. Whereas the $12$ time components of such as fields $(e_{0}^{a}, w_{0}^{a},l_{0}^{a}, A_{0}^{a})$ are Lagrange multipliers for the $12$ primary constraints $(\Phi^{a}_{1},\Phi^{a}_{2},\Phi^{a}_{3},\Phi^{a}_{4})$, out of which only $6$ (linear combinations) must be first-class ones, corresponding to the 6 gauge symmetries (3 overall Lorentz transformations plus 3 overall diffeomorphism) of the theory. Under the conditions spelled above, we have in total 14 constraints, 6 of them must be first-class, and the remaining 8 must be second-class. The Lagrange multipliers do not contribute to the degrees of freedom count. Therefore, we are now in a position to count the physical phase-space dimensionality $\mathcal{N}$ of our theory from the following formula:
\begin{equation}
\mathcal{N}=\mathcal{P}-2\times\mathcal{F}-\mathcal{S},\label{DoF}
\end{equation}
where $\mathcal{P}$ is number of dynamical variables, $\mathcal{F}$ is the number of first-class constraints, and $\mathcal{S}$ is the number of second-class constraints. Hence, for the present case, this allows us to conclude that the dimension of the physical phase space  is
\begin{equation}
\mathcal{N}=24-6\times2-8=4.
\end{equation}
Thus,   when $l^{\mu}_{d}$ is invertible and $\beta_{3}$ is zero, general exotic bi-gravity has $2$ physical degrees of freedom, as expected.

Notice that the analysis in the case \textit{(ii)}, when $\beta_{4}$ is zero and $e^{\mu}_{a}$ is invertible, is similar to the case \textit{(i)}.  Hence, the number of physical degrees of freedom is also $2$. In conclusion, in both cases, there are two physical degrees of freedom.

On the other hand, for generic (non-zero) values of the coupling constants $\beta_{n}$, the theory has three secondary constraints: $\Theta^{0}$, $\Gamma^{0}$, $\Psi^{0}$. Therefore, the number of constraints has increased by one. This implies that there are 15 constraints; out of which 6 must be first-class ones and 9 second-class ones. This suggests that the dimension of the physical phase-space turns out to be
\begin{equation}
\mathcal{N}=24-6\times2-9=3!,
\end{equation}
which leads to an inconsistency in the dimensionality of the phase space. From this we conclude that the theory only has two equivalent sectors, where there are two physical degrees of freedom. In this setting, the model has the same number of physical degrees of freedom as ZDG.

Finally, we may analyze the case in which the parameters $\beta_{1}$, $\beta_{3}$, and $\gamma_{1}$, are going to zero. At the action level, the interaction between the two dreibein only manifests itself through the constants $\beta_{1}$, $\beta_{3}$, and $\gamma_{1}$, so that, in this case, the dreibein field $e_{I}^{a}$ is effectively decoupled from the other one $l_{I}^{a}$, requiring only invertibility of $l_{I}^{a}$ to derive the necessary secondary constraints.  Using our constraints analysis, we can infer that from all the primary constraints (\ref{C-P2}), the ones that remain are $\Phi_{1}^{a}\mid_{\beta_{1}=0}$, $\Phi_{3}^{a}\mid_{\beta_{3}=0}$, and $\Phi_{4}^{a}$, whereas, from the integrability conditions Eqs. (\ref{Int1})-(\ref{Int3}), we can only derive one secondary constraint, namely, $\Theta^{0}$. In total, there are now 10 constraints, out of which 6 are first-class and  4 second-class. Hence,  according to Eq. (\ref{DoF}), the resulting theory,  given by the following action
\begin{eqnarray}
S[w^{a}_{\alpha},l^{a}_{\alpha},A^{a}_{\alpha}]&=&\int d^{3}x\,\varepsilon^{\mu\nu\gamma}\left[\frac{1}{2}\alpha_{2}\left(A_{a\mu}\partial_{\nu}A_{\gamma}^{a}+\frac{1}{3}\epsilon^{abc}A_{a\mu}A_{b\nu}A_{c\gamma}\right)+\beta_{2}l_{a\mu}\nabla_{\nu}l^{a}_{\gamma}\right.\nonumber\\
&&\left.+\frac{1}{2}\alpha_{1}\left(w_{a\mu}\partial_{\nu}w_{\gamma}^{a}+\frac{1}{3}\epsilon^{abc}w_{a\mu}w_{b\nu}w_{c\gamma}\right)+\beta_{4}l_{a\mu}D_{\nu}l^{a}_{\gamma}+\gamma_{2}l_{\mu}^{a}l_{\nu}^{b}l_{\gamma}^{c}\right],
\end{eqnarray}
has a physical phase space of dimension $2$, or $1$ physical degree of freedom, as  was reported in Ref. \cite{GEMG}.
\section{ CONCLUSIONS}
\label{conclusions}
In this paper, we studied the dynamical structure of general exotic bi-gravity theory in the context of Faddeev-Jackiw Hamiltonian formulation. To begin with, we performed a space-time decomposition of fields and wrote the action in its first-order form, i.e., linear in the velocities (\ref{action}). After that, we constructed the proper pre-symplectic matrix (\ref{sym1}), whose zero-modes allowed us to derive the set of primary constraints (\ref{C-P2}). As shown in subsection \ref{IntCondASec}, by combining equations of motion with consistency conditions for the primary constraints, we obtained a new matrix, which is not a square one, but still has linearly independent zero-modes. These zero-modes generated a set of integrability conditions (\ref{Int1})-(\ref{Int3}) when multiplied by the gradient of the Hamiltonian. Subsequently, these integrability conditions allowed us to extract proper secondary constraints responsible for completely removing the Boulware-Deser ghosts and thus rendering the model consistent. Along the same lines, we exploited the consistency of the secondary constraints and showed the absence of tertiary and further constraints, but derived three scalar equations (\ref{71})–(\ref{72}) establishing the most general relationship between all the parameters and fields defining the action (\ref{principle}) of the model. We then used such expressions to study the AdS background.

Once all the constraints had been found out, we incorporated them into the kinetic part of the action in Eq. (\ref{action}) to construct a new one (\ref{new}), nevertheless resulting in a singular  matrix. So, this pre-symplectic matrix contains all the information on the existing symmetries of the theory. At this point, the zero-modes played a very important role. We observed that only the zero-modes that turn out to be orthogonal to the Hamiltonian gradient, represent the canonical generators of the gauge symmetries of the model. Considering these zero-modes, we explicitly derived the correct gauge transformations of the canonical variables on the constraints surface (\ref{gauge1})-(\ref{gauge4}). Further, we successfully recovered the diffeomorphism symmetry by mapping the gauge parameters appropriately.

Finally, the correct identification of all the constraints, in turn, allowed us to revise the counting of physical degrees of freedom. We have found that, by setting to zero one of the four coupling constants $\beta_{n}$, and assuming the invertibility of some of the dreibeine, the general exotic bi-gravity theory has two degrees of freedom, as expected. Our work suggests that the Faddeev-Jackiw symplectic approach, provides a more economical way in deriving the physical constraints and the gauge structure for massive gravity  and bi-gravity models where primary, secondary, tertiary and quartic constraints could be present.

\section*{ACKNOWLEDGMENTS}
We acknowledge partial support from Sistema Nacional de Investigadores (Mexico) and Direcci\'on de Apoyo a Docentes, Investigaci\'on y Posgrado de la Universidad de Sonora.

\end{document}